\documentclass{aastex631}

\usepackage{listings}
\submitjournal{ApJ}

\shorttitle{M3 Tidal Stream}
\shortauthors{Yang et al.}
\graphicspath{{./}{figures/}}

\begin{document}

\title{The Spectacular Tidal Tails of Globular Cluster M3 (NGC 5272)}

\correspondingauthor{Jing-Kun Zhao}
\email{zjk@bao.ac.cn}

\author[0000-0001-7609-1947]{Yong Yang}
\affiliation{CAS Key Laboratory of Optical Astronomy, National Astronomical Observatories, Chinese Academy of Sciences, Beijing 100101, People's Republic of China}
\affiliation{School of Astronomy and Space Science, University of Chinese Academy of Sciences, Beijing 100049, People's Republic of China},
\author[0000-0003-2868-8276]{Jing-Kun Zhao}
\affiliation{CAS Key Laboratory of Optical Astronomy, National Astronomical Observatories, Chinese Academy of Sciences, Beijing 100101, People's Republic of China},
\author{Xin-Zhe Tang}
\affiliation{CAS Key Laboratory of Optical Astronomy, National Astronomical Observatories, Chinese Academy of Sciences, Beijing 100101, People's Republic of China}
\affiliation{Department of Physics, College of Science, Tibet University, Lhasa 850000, People's Republic of China},
\author[0000-0002-5805-8112]{Xian-Hao Ye}
\affiliation{CAS Key Laboratory of Optical Astronomy, National Astronomical Observatories, Chinese Academy of Sciences, Beijing 100101, People's Republic of China},
\author[0000-0002-8980-945X]{Gang Zhao}
\affiliation{CAS Key Laboratory of Optical Astronomy, National Astronomical Observatories, Chinese Academy of Sciences, Beijing 100101, People's Republic of China}
\affiliation{School of Astronomy and Space Science, University of Chinese Academy of Sciences, Beijing 100049, People's Republic of China}



\begin{abstract}

We provide a detailed analysis on tidal tails of the globular cluster M3 (NGC 5272). We first discover clear extra-tidal structures with slight S-shape near the cluster. This inspires us to examine the existence of its long tidal tails. We highlight potential stream stars using proper motions (PMs) of a model stream combined with the cluster's locus in a color-magnitude diagram (CMD). A 35$\degr$ long leading tail and a 21$\degr$ long trailing tail are successfully detected at the same time. Their corresponding overdensities can be recognized in CMD and PM space after subtracting background. We estimate stream width, star number density and surface brightness for both tails, as well as the distance variation along the entire stream. We then verify the connection of M3 and the Sv\"{o}l stream. Finally, we tabulate 11 member stars belonging to the M3 tidal stream with available spectroscopic observations.

\end{abstract}



\section{Introduction} \label{sec:intro}

The Milky Way is our host galaxy and the only one galaxy we can learn about in great detail. Understanding its structure and evolution has been one of the most important topics in astronomy for the past decades. Specifically, studying the Galactic disk reveals kinematic substructures \citep[e.g.,][]{2009ApJ...692L.113Z,2017ApJ...844..152L,2018Natur.561..360A,2018ApJ...868..105Z,2021ApJ...922..105Y}, streams embedded in the disk \citep[e.g.,][]{2020NatAs...4.1078N,2021ApJ...907L..16R,2021AJ....162..171Y}, dynamic connections between inner and outer disk \citep[e.g.,][]{2021ApJ...919...52Z,2022ApJ...936L...7C}, imprints from merger events \citep[e.g.,][]{2021SCPMA..6439562Z,2023MNRAS.520.1913W}, and so on. Focusing on the Galactic halo disentangles plentiful stellar streams \citep[e.g.,][]{2016ASSL..420...87G,2018ApJ...862..114S,2020ApJ...902...89T,2020ApJ...904...61Z,2021ApJ...914..123I,2022ApJ...935L..38Y,2023ApJ...945L...5Y}, merger history \citep[e.g.,][]{1994Natur.370..194I,2018Natur.563...85H,2022ApJ...926..107M}, dark matter mass \citep[e.g.,][]{2008ApJ...684.1143X,2022MNRAS.516..731B}, etc. In this work, we concentrate on stellar streams, as one type of important materials probing the Milky Way's structure and evolution.

Stellar stream is not an unfamiliar concept. Before Gaia era, photometric data (e.g., SDSS) were widely used to look for streams by adopting a matched filter analysis \citep{2002AJ....124..349R}, and plenty of streams were found \citep[e.g.,][]{2006ApJ...643L..17G,2009ApJ...693.1118G,2012ApJ...760L...6B,2014MNRAS.442L..85K,2018ApJ...862..114S}. The technique relies on a truth that stars belonging to a stream form a particular stellar population such that they can be well described with an isochrone in a color-magnitude diagram (CMD). The isochrone filter is placed at various distance modulus to account for stellar population at different distances. If there are any stream stars, they will be highlighted and noticed in celestial coordinate plane. With arrival of Gaia especially after data release 2 \citep{2018A&A...616A...1G}, high precision astrometry is available and combined with photometry to detect streams. A revolutionary achievement is development of \texttt{STREAMFINDER} \citep{2018MNRAS.477.4063M}, which successfully found dozens of new streams and dramatically enlarged stream family \citep{2018MNRAS.481.3442M,2021ApJ...914..123I}. Besides, an ingenious modified matched filter was introduced by \citet{2019ApJ...884..174G}, where in addition to a traditional isochrone filter, proper motion (PM) filters were created by computing orbits for a cluster to predict what PMs as functions of position would look like if the cluster had tidal tails. The method is a result of making use of Gaia and good at detecting streams of rather weak signals \citep[e.g.,][]{2022ApJ...929...89G,2022A&A...667A..37Y}. 

In this work, we apply matched filter methods to study tidal tails of the globular cluster M3 (NGC 5272), for which there are already some implications of its tidal disruption \citep[e.g.,][]{2021ApJ...909L..26B,2022MNRAS.513.1958W}. Section~\ref{sec:data} introduces the data. Section~\ref{sec:M3_nearby} describes extra-tidal features in nearby area of M3. Section~\ref{sec:stream} characterizes the whole stellar stream. A conclusion is given in Section~\ref{sec:summary}. 

\section{Gaia data} \label{sec:data}

We take advantage of improved astrometry and photometry from Gaia data release 3 \citep[DR3,][]{2021A&A...649A...1G,2021A&A...649A...2L,2021A&A...649A...3R}. Stars nearby M3 within sky box $180\degr <$ R.A. $< 280\degr$ and $0\degr <$ Dec. $< 40\degr$ are retrieved. Galactic latitude $b > 15\degr$ is further required to get rid of most disk stars. In order to ensure good astrometric and photometric solutions, only stars with a renormalized unit weight error (RUWE) $<$ 1.4 and $|C^*| < 3\sigma_{C^*}$ are retained, where $C^*$ is the corrected BP and RP flux excess factor that was introduced by \citep{2021A&A...649A...3R} to identify sources for which the $G$–band photometry and BP and RP photometry are not consistent. All major GCs in this field are then masked using tidal radius ($r_t$) from \citet[][2010 edition]{1996AJ....112.1487H}, including M3 itself ($r_t = 28.7'$), NGC 5466 ($r_t = 15.68'$), NGC 4147 ($r_t = 6.08'$), NGC 5024 ($r_t = 18.37'$), NGC 5053 ($r_t = 11.43'$) and NGC 6205 ($r_t = 21'$). 

All stars are extinction-corrected using the \citet{1998ApJ...500..525S} maps as re-calibrated by \citet{2011ApJ...737..103S} with $RV$ = 3.1, assuming $A_{G}/A_{V} = 0.83627$, $A_{BP}/A_{V} = 1.08337$, $A_{RP}/A_{V} = 0.63439$\footnote{These extinction ratios are listed on the Padova model site \url{http://stev.oapd.inaf.it/cgi-bin/cmd}.}. We also calculate magnitude errors in BP and RP bands $\sigma_{BP}$ and $\sigma_{RP}$ using a propagation of flux errors (see CDS website\footnote{\url{https://vizier.u-strasbg.fr/viz-bin/VizieR-n?-source=METAnot&catid=1350&notid=63&-out=text}.}), and obtain color $BP - RP$ and corresponding error $\sigma_{color}$ simply through $\sqrt{\sigma^2_{BP} + \sigma^2_{RP}}$.

\section{Tidal features around M3} \label{sec:M3_nearby}

\subsection{A revised matched filter}

We first have a look at the surrounding of M3 to figure out whether there are any tidal features around it. We introduce a revised matched filter\footnote{To avoid confusion, we refer to this method as "revised" matched filter, distinguished from the "modified" matched filter \citep{2019ApJ...884..174G} used in Section~\ref{sec:stream}.} method that highlights cluster-like stars in CMD as well as in PM space.

The traditional matched filter \citep{2002AJ....124..349R} is based on photometric data. Weights are obtained from ratios of CMD densities of the target population to the background, and stars are assigned different weights according to their positions in CMD. By doing so, stars in regions of CMD with high contrast between the target and background will be weighted heavily. Following this thought, and under a reasonable assumption that extra-tidal stars located not too far from the cluster have similar PMs to the cluster, we consider that stars can be further weighted in PM space in the same way. 

\begin{figure*}
	\includegraphics[width=\linewidth]{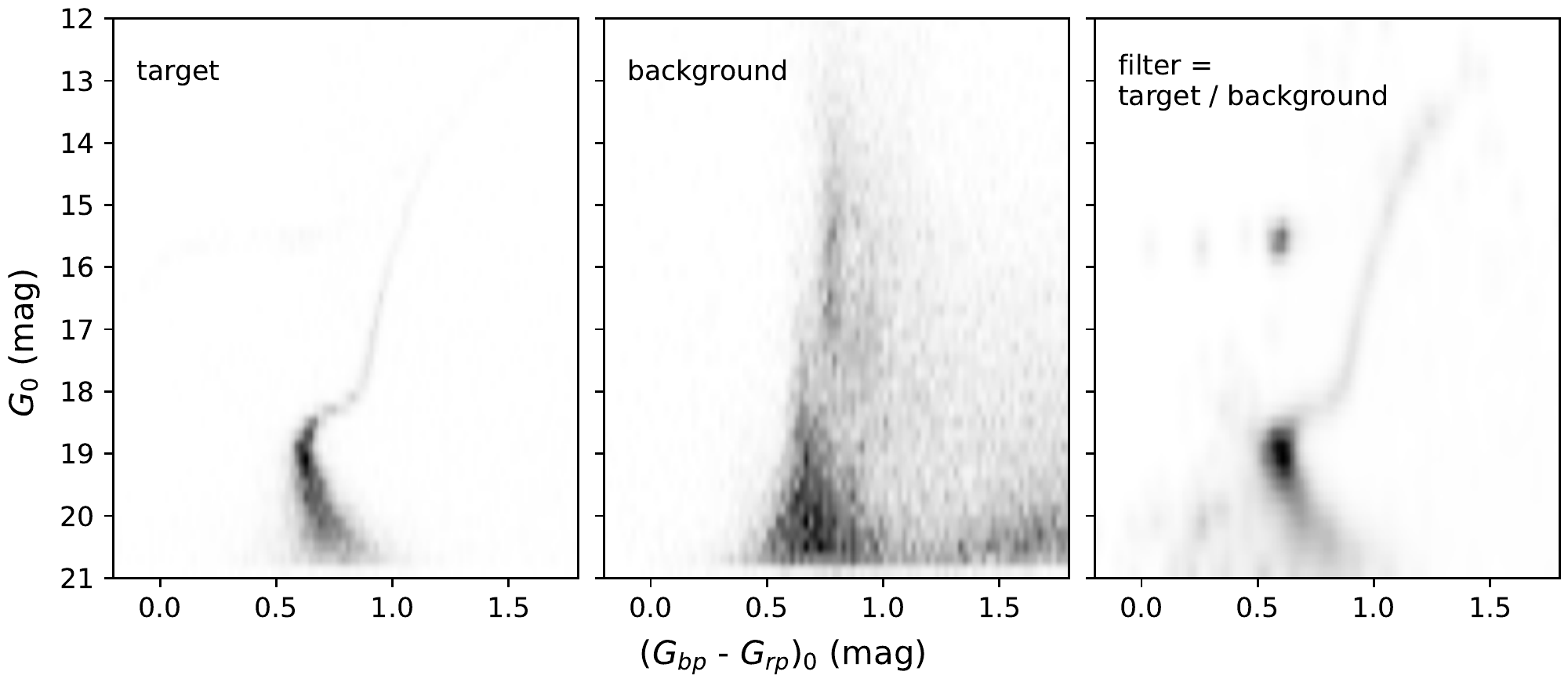}
	\includegraphics[width=\linewidth]{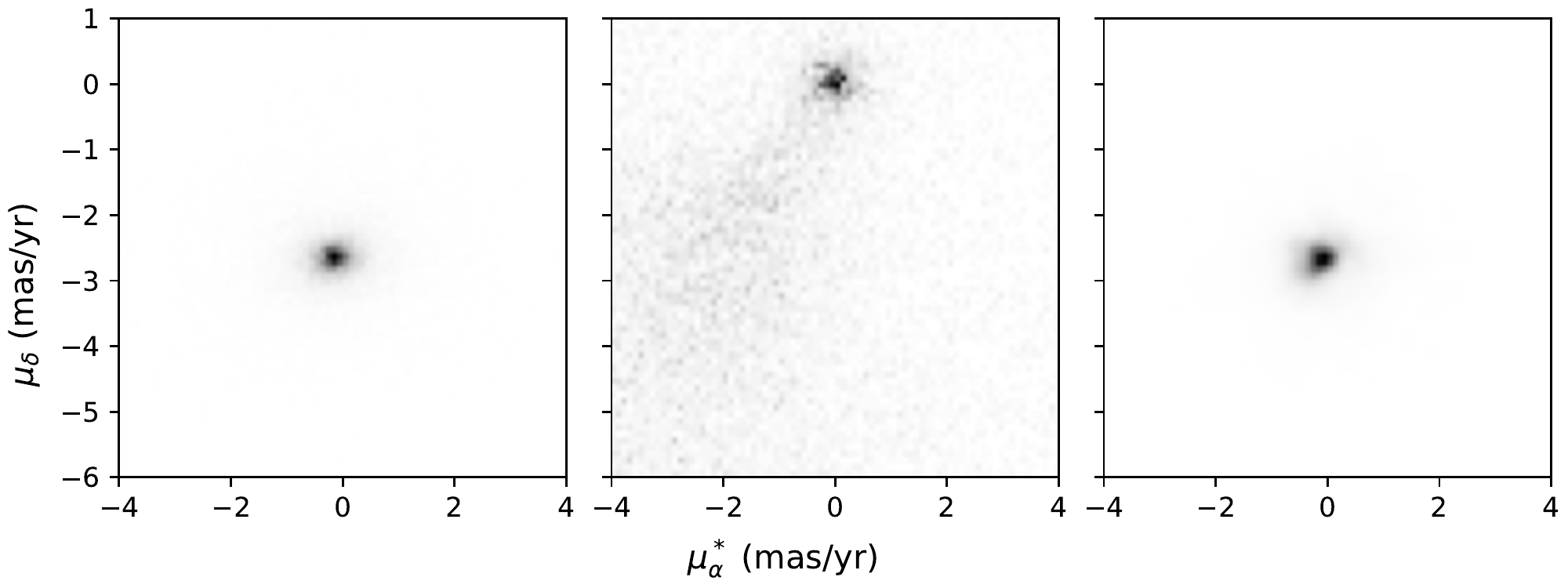}
	\caption{The upper and lower panels show the construction of the filters in CMD and PM space. From left to right columns are distributions (linear stretch) of target, background and filters, respectively. The CMD bin size is 0.02 mag in color and 0.2 mag in $G_0$ magnitude. The PM bins are both 0.1 mas yr$^{-1}$. Filters are smoothed with a Gaussian kernel of $\sigma$ = 1 pixel. 
		\label{fig:new_mf}}
\end{figure*}

We select Gaia data nearby M3 in $203\degr <$ R.A. $< 208\degr$ and $24\degr <$ Dec. $< 33\degr$. M3 stars within the tidal radius are the target stellar population and the rest of stars between $24\degr <$ Dec. $< 27\degr$ and $30\degr <$ Dec. $< 33\degr$ act as the background. The left and center columns of Figure~\ref{fig:new_mf} show the target and background densities in CMD and PM space. Two filters are created by dividing the target by the background and further smoothed, as displayed in the right column in Figure~\ref{fig:new_mf}. The filter of CMD gives higher weights to the main-sequence turnoff and a portion of the horizontal branch due to that fewer field stars are as this blue in color. The filter in PM space highlights stars whose PMs are close to M3. Two kinds of weights of stars assigned from the filters are multiplied as the final weights. 

\subsection{Results}

\begin{figure}
	\includegraphics[width=0.5\linewidth]{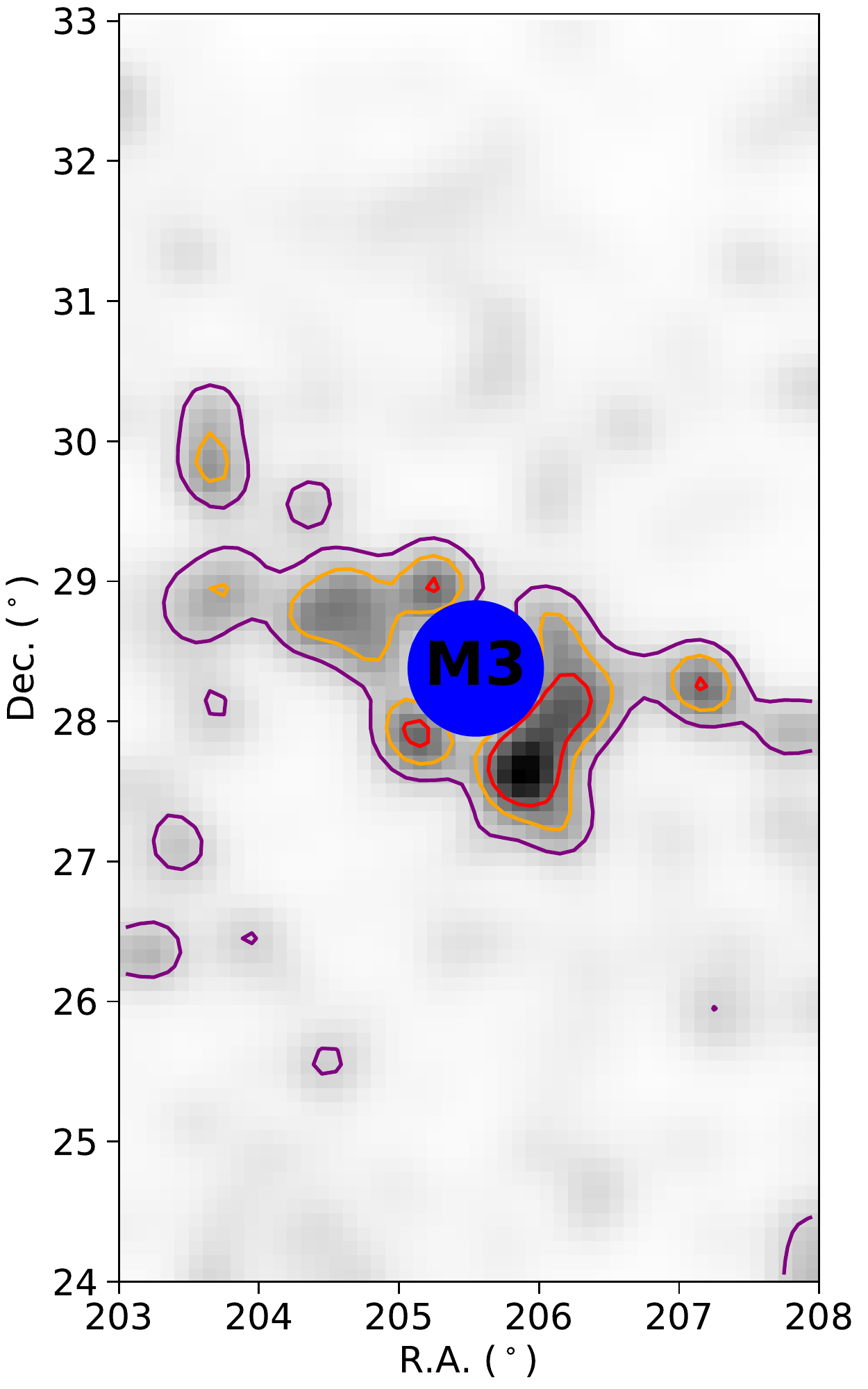}
	\caption{Weighted map with linear stretch of M3 surrounding area. The pixel size is $0.1\degr \times 0.1\degr$ and the map is smoothed with a Gaussian kernel of $\sigma$ = 0.2$\degr$. The contours represent an S/N of 3 (purple), 6 (orange), and 9 (red). The blue circle indicates the M3 tidal radius. 
		\label{fig:extratidal}}
\end{figure}

The final weights of stars are summed in sky pixels and a smoothed weighted map of the surrounding of M3 can be produced as shown in Figure~\ref{fig:extratidal}. The contours represent a signal-to-noise ratio (S/N) of 3 (purple), 6 (orange), and 9 (red), where S/N is estimated through a formula = (smoothed map - mean of smoothed background) / standard deviation of smoothed background. The blue circle indicates the M3 tidal radius. It can be seen that some extra-tidal structures are revealed clearly here, including slight S-shape which is connected to the mass-loss phenomenon of globular clusters. This means that there is tidal disruption in the surrounding area of M3, and it motivates us to investigate whether the cluster has long tidal tails farther away. 

\section{Tidal tails of M3} \label{sec:stream}

\subsection{A modified matched filter}

\citet{2019ApJ...884..174G} introduced a modified matched filter to detect stellar streams of globular clusters by making use of Gaia photometry and astrometry. The way it assigns weights to stars is different from that of the revised matched filter mentioned before. Not using a ratio of the target to the background, the modified matched filter adopts a Gaussian error distribution to generate a star's weight. The method compare how close the star is to a cluster isochrone in CMD by taking into account photometric uncertainties. It further considers how similar the star's PMs are to the cluster's orbit, which is an approximation of tidal tails, with astrometric uncertainties included. 

\begin{figure*}
	\includegraphics[width=0.36\linewidth]{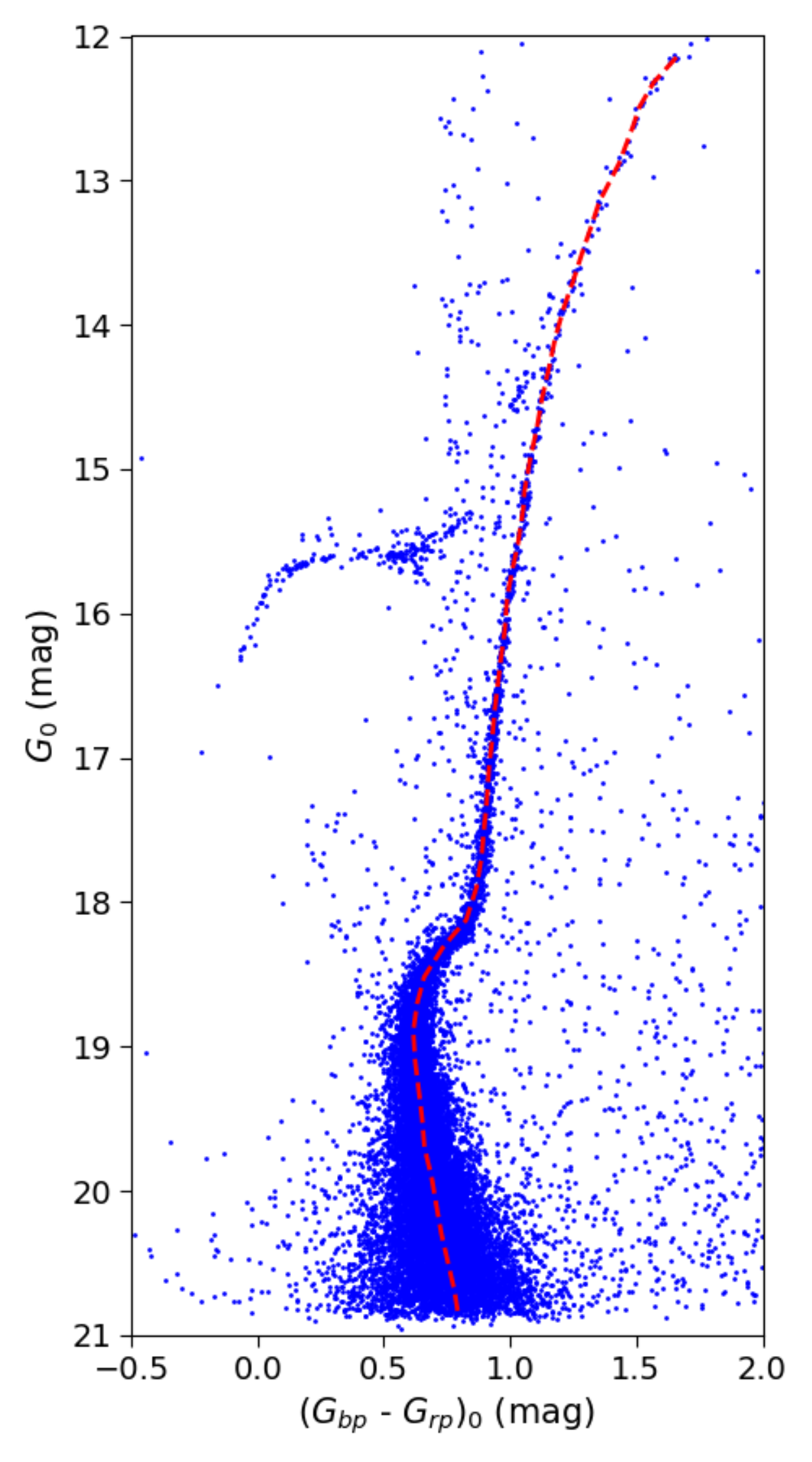}
	\includegraphics[width=0.64\linewidth]{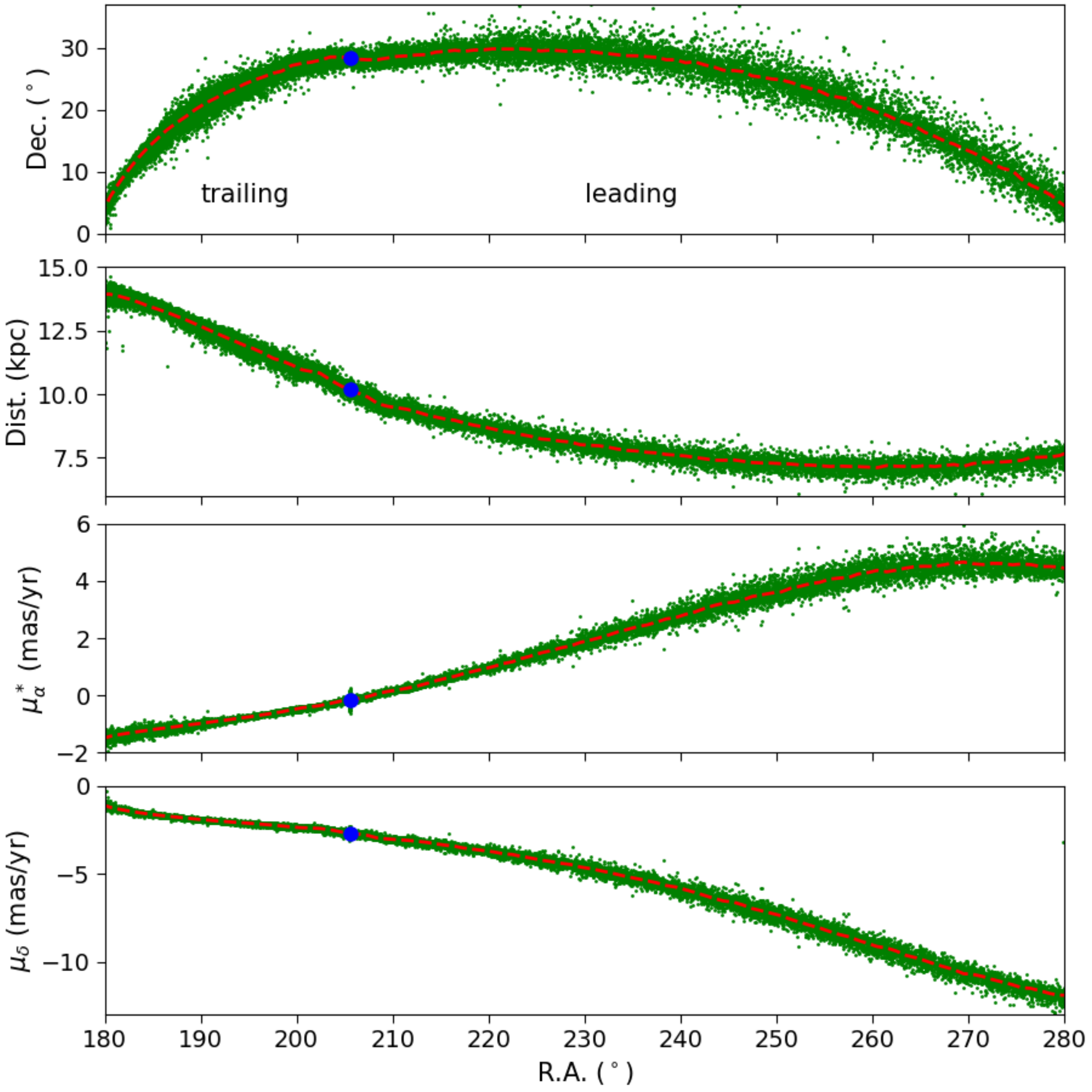}
	\caption{The left panel is a CMD of those stars within M3's tidal radius shown in blue dots. The red dashed line is the locus of the red giant branch and main sequence, obtained by calculating median values along $G_0$. The right panel shows the model stream (green dots) in planes of Dec., heliocentric distance, $\mu^*_{\alpha}$ and $\mu_{\delta}$ as a function of R.A., from top to bottom. The red dashed lines represent computed medians along R.A.. The blue circle stands for M3. 
		\label{fig:filters}}
\end{figure*}

To get a CMD filter, we retrieve stars located within M3's tidal radius as shown with blue dots in the left panel of Figure~\ref{fig:filters}. For stars of red giant branch and main sequence, we compute their color medians along $G_0$ magnitude (red dashed line). This serves as the locus of M3 on CMD and we call it CMD-filter later. 

To get an estimate of tidal tails for M3, we model its disruption by a Lagrange-point stripping method \citep{2014MNRAS.445.3788G,2019MNRAS.487.2685E} under a static Galactic potential in the presence of a moving Large Magellanic Cloud (LMC). The potential consists of a halo taken from \citet{2017MNRAS.465...76M} and non-halo components from \citet[model I]{2017A&A...598A..66P}. In detail, there is a Plummer bulge \citep{1911MNRAS..71..460P} with scale radius = 0.3 kpc and mass = $1.0672\times10^{10} M_{\sun}$, a Miyamoto–Nagai \citep{1975PASJ...27..533M} thin disk with scale length = 5.3 kpc, scale height = 0.25 kpc, and mass = $3.944\times10^{10} M_{\sun}$, a Miyamoto–Nagai thick disk with scale length = 2.6 kpc, scale height = 0.8 kpc, and mass = $3.944\times10^{10} M_{\sun}$, and a NFW halo \citep{1996ApJ...462..563N} with virial mass = $1.37\times10^{12} M_{\sun}$ and concentration = 15.4. LMC is modeled by a Hernquist potential \citep{1990ApJ...356..359H} with scale radius = 17.13 kpc and mass = $1.5\times10^{11} M_{\sun}$ \citep{2022ApJ...928...30L}. LMC's position and velocity are from \citet{2018A&A...616A..12G}. M3 is modeled as a Plummer sphere where its mass is $4.09\times10^5 M_{\sun}$ and scale radius is 5.47 pc \citep{2018MNRAS.478.1520B}. The cluster's kinematic parameters are from \citet{2021MNRAS.505.5978V} and \citet[][2010 edition]{1996AJ....112.1487H}. Here the position of the sun is set as $(R_{\sun}, Z_{\sun})$ = (8.122, 0.0208) kpc \citep{2018A&A...615L..15G,2019MNRAS.482.1417B}, and the solar velocities are set to $(V_{R,\sun}, V_{\phi,\sun}, V_{Z,\sun})$ = (-12.9, 245.6, 7.78) km s$^{-1}$ \citep{2018RNAAS...2..210D}, respectively. After those parameters are settled, M3 is initialized 2 Gyr ago and integrated forward from then on until now, releasing particles at Lagrange points in both leading and trailing directions. Finally we obtain a model stream for the cluster through this way. More details of this process can be found in \citet{2022A&A...667A..37Y,2022MNRAS.513..853Y}.

We present the model stream in the right panel of Figure~\ref{fig:filters} shown with green dots. It is interesting that the mock tails on two sides of M3 also have slight S-shape just similar to observed one as revealed in Section~\ref{sec:M3_nearby}. Given the stream particles, we calculate their medians along R.A. and show them with red dashed lines. Hereafter we refer to median lines in two PM planes as PM-filters.

On CMD, individual stars are assigned weights according to their color differences from the CMD-filter, assuming a Gaussian error distribution:
\begin{equation}
	w_{\rm CMD} = \frac{1}{\sqrt{2\pi} \sigma_{color}} {\rm exp}
	\left[ -\frac{1}{2} \left(\frac{color - color_0}{\sigma_{color}} \right)^2  \right]  .
\end{equation}
Here $color$ and $\sigma_{color}$ denote $BP - RP$ and corresponding errors. $color_0$ is determined by the CMD-filter at a given $G_0$ magnitude of a star. 

Using PM-filters, a star's weight is computed as:
\begin{equation}
	w_{\rm PMs} = \frac{1}{2\pi \sigma_{\mu^*_{\alpha}}\sigma_{\mu_{\delta}}} {\rm exp}
	\left\lbrace -\frac{1}{2} \left[ \left( \frac{\mu^*_{\alpha} - \mu^*_{\alpha,0}}{\sigma_{\mu^*_{\alpha}}} \right)^2 +
	\left( \frac{\mu_{\delta} - \mu_{\delta,0}}{\sigma_{\mu_{\delta}}} \right)^2 \right] \right\rbrace .
\end{equation}
Here $\mu^*_{\alpha}$, $\mu_{\delta}$, $\sigma_{\mu^*_{\alpha}}$ and $\sigma_{\mu_{\delta}}$ are measured PMs and corresponding errors of stars. $\mu^*_{\alpha,0}$ and $\mu_{\delta,0}$ are components of PMs predicted at each star's R.A. based on the PM-filters. 
 
The final weights are obtained by multiplying both $w_{\rm CMD} \times w_{\rm PMs}$, and stars' weights are summed in sky pixels to expose possible structures. Considering a distance range covered by the mock stream (Dist. plane in Figure~\ref{fig:filters}), three types of $w_{\rm CMD}$ are computed by placing CMD-filter at current position (M3's Dist. = 10.18 kpc), upward to brighter side by 0.5 mag (leading tail Dist. = 8 kpc), and downward to fainter side by 0.6 mag (trailing tail Dist. = 13.5 kpc). They are coadded together and multiplied by $w_{\rm PMs}$ to account for the whole stream spanning a large extent of distance.

\subsection{A whole look} \label{sec:awholelook}

\begin{figure*}
	\includegraphics[width=\linewidth]{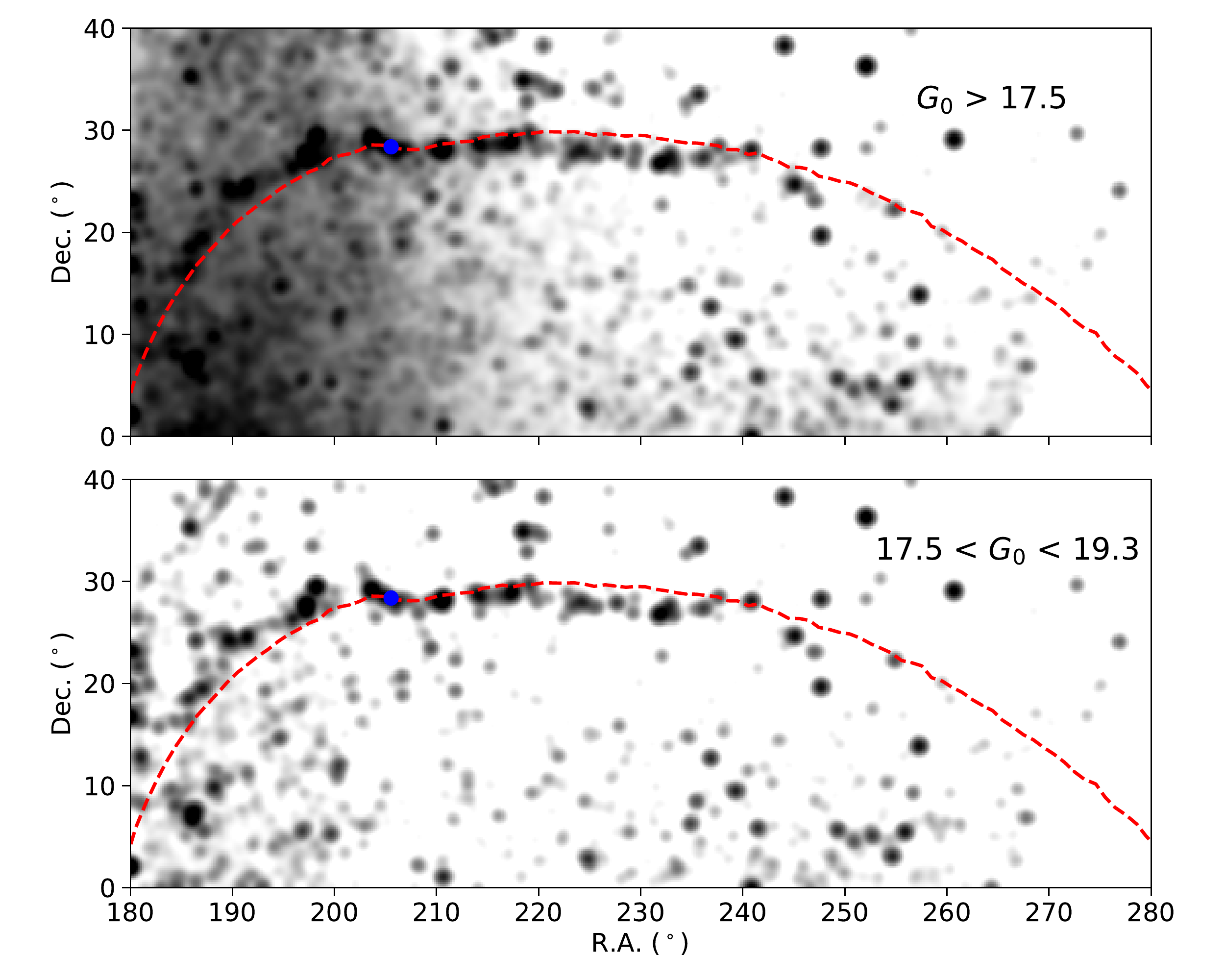}
	\caption{Top: Log stretch of a weighted map. The sky pixel width is 0.2$\degr$ and the map is smoothed with a Gaussian kernel of $\sigma$ = 0.5$\degr$. Only stars fainter than 17.5 mag are plotted. The blue circle and red dashed line represent M3 and median positions of the model stream as displayed in Figure~\ref{fig:filters}. Bottom: similar to the top one but only stars between 17.5 and 19.3 mag are plotted.
		\label{fig:weighted_map}}
\end{figure*}

With all stars' weights, we create a weighted map of sky and show it in the upper panel of Figure~\ref{fig:weighted_map}. The lack of data in right bottom corner is due to that we discard stars with $b < 15\degr$ as mentioned before. We also mark M3 with a blue circle and show median positions of the mock stream with the red dashed line which is the same as the first panel in Figure~\ref{fig:filters}. Extending from the cluster, the leading tail is clearly exposed that almost follows the mock one as far as R.A. = 248$\degr$. The trailing tail, however, is very ambiguous and highly contaminated by field stars. The upper panel only shows stars fainter than $G_0$ = 17.5 mag (corresponding to beginning of the leading tail's main sequence turn-off, see Figure~\ref{fig:leading_CMD_PM}) because random brighter field stars might introduce large noises due to their rather low photometric and astrometric uncertainties, although there are still several noises appearing on leading side which do not represent actual physical overdensities. 

To distinguish the trailing tail from field stars, we further cut off stars fainter than $G_0$ = 19.3 mag. The ambiguity is mainly as a result of that stream and non-stream stars have similar PMs and distributions in CMD, and hence they are both given high weights. However, most of field stars have $G_0 >$ 19.3 mag and brighter ones hardly overlap with the cluster in CMD. Therefore, restriction of $G_0 <$ 19.3 mag is able to remove a majority of contaminants (this is elaborated in Appendix~\ref{appendix}). In the second panel of Figure~\ref{fig:weighted_map}, only stars satisfying 17.5 mag$< G_0 <$ 19.3 mag are plotted. In spite of being a bit noisy, the trailing tail becomes evident than before and follows the model as well.

\begin{figure*}
	\includegraphics[width=\linewidth]{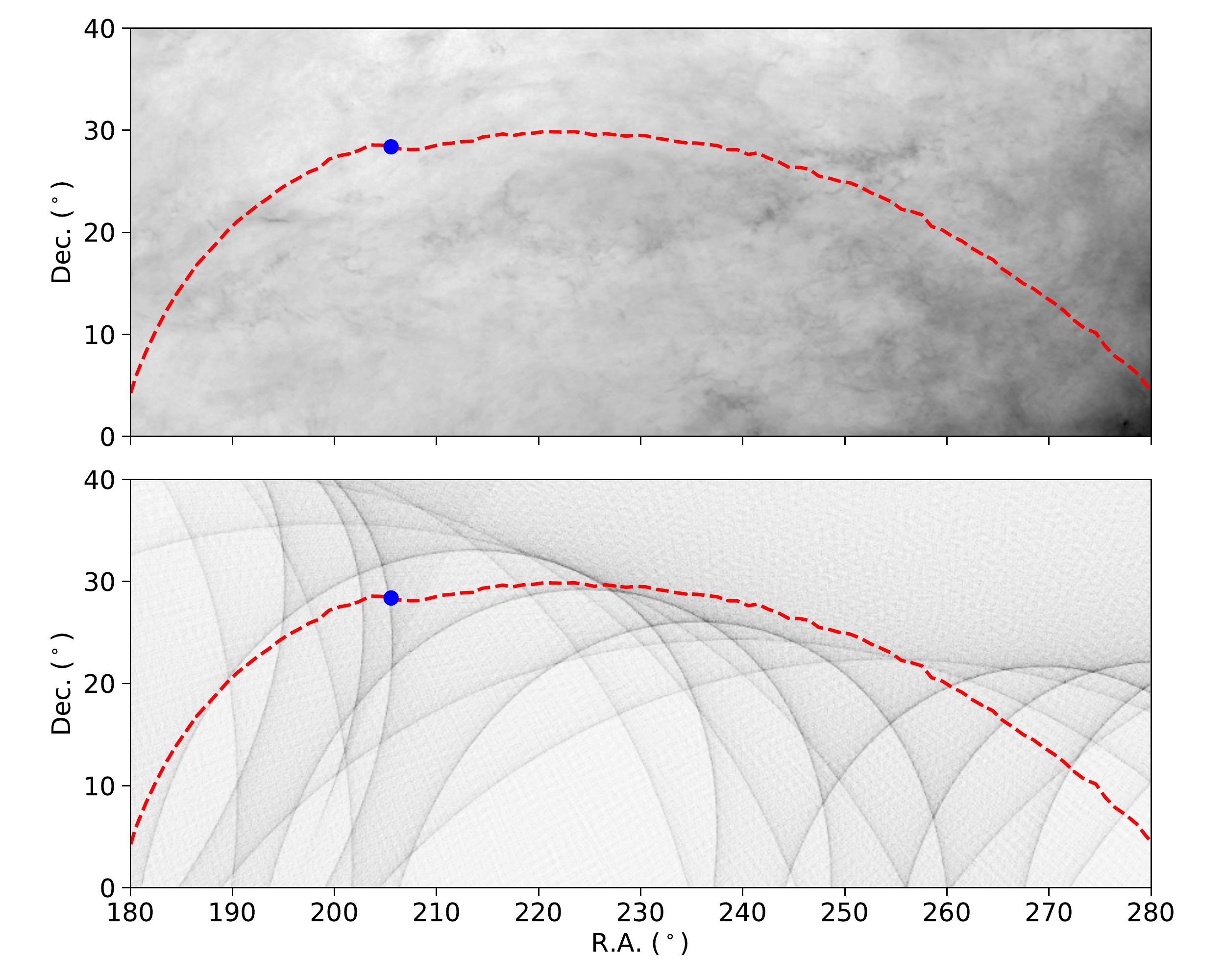}
	\caption{The top and bottom panels present the dust extinction map extracted from \citet{1998ApJ...500..525S} and $Gaia$'s scanning pattern covered by the DR3, respectively. The blue circle and red dashed line mean the same as Figure~\ref{fig:weighted_map}.
		\label{fig:sfd_scan}}
\end{figure*}

Figure~\ref{fig:sfd_scan} presents the dust extinction map extracted from \citet{1998ApJ...500..525S} and $Gaia$'s scanning pattern covered by the DR3, with higher values represented by darker colors. The model stream's locus is still plotted as a reference. No major structures of the extinction exist because this field is around the Galactic north pole. The stream does not strictly follow the scanning tracks either. If signals in Figure~\ref{fig:weighted_map} are caused by Gaia's observing mode, we should recover several impressions same as this scanning pattern. Consequently, we confirm that detected signals are not fake.

\subsection{The leading tail}

\begin{figure*}
	\includegraphics[width=\linewidth]{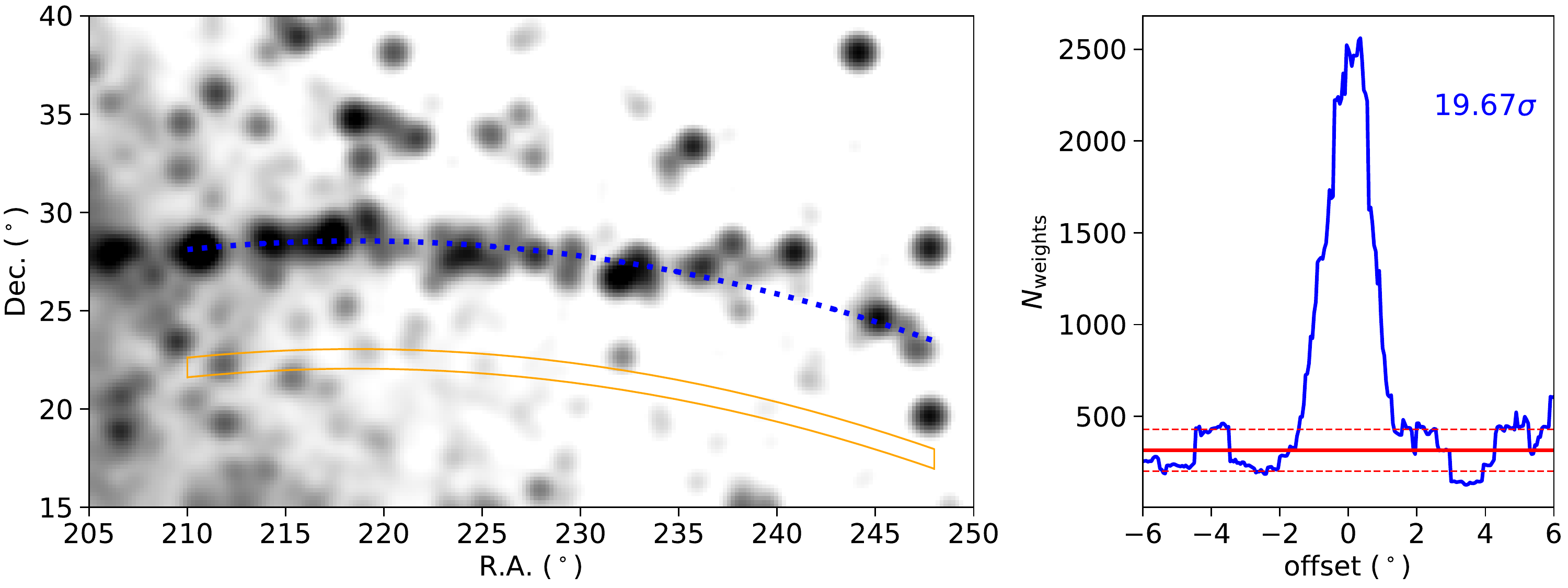}
	\caption{Left: a zoom-in look at the leading tail. Here the blue dotted line is a second-order polynomial fit to the trajectory. The area enclosed by orange solid lines is a mask parallel to the fit used to create the lateral profile; Right: the lateral profile of the leading tail. Red straight lines indicate average level of background and 1$\sigma$ extent.
		\label{fig:leading_tail}}
\end{figure*}

We zoom in the leading tail in the left panel of Figure~\ref{fig:leading_tail}. Since it is almost horizontal in R.A.-Dec. plane, we do not further rotate to a stream-aligned coordinate here. The blue dotted line means a second-order polynomial fit to the trajectory. The polynomial expression is
\begin{equation}
	\rm{Dec.} = -5.89976403 \times 10^{-3} \times \rm{R.A.}^2 + 2.57955221 \times \rm{R.A.} -2.53417384 \times 10^2
	\label{equation3}
\end{equation}
with 210$\degr$ $<$ R.A. $<$ 248$\degr$. Below the leading tail is a mask shown with orange solid lines. Its width is 1$\degr$ and long side is parallel to the polynomial. To estimate the stream's width, we move the mask across the stream from bottom to top and sum all weights of stars fainter than $G_0$ = 17.5 mag in the mask (again, to reduce large noises caused by bright stars), to create a one-dimensional stream profile as shown with the blue solid line in the right panel of Figure~\ref{fig:leading_tail}. This is directly analogous to the T statistic of \citet{2009ApJ...693.1118G}. The stream is almost enclosed within |offset| $< 1.5\degr$, and its peak is 19.67$\sigma$ above the background noise outside this range. Based on the profile, we find an estimate of its width (full width at half maximum, FWHM) to be $\sim 1.6\degr$. 

\begin{figure*}
	\includegraphics[width=\linewidth]{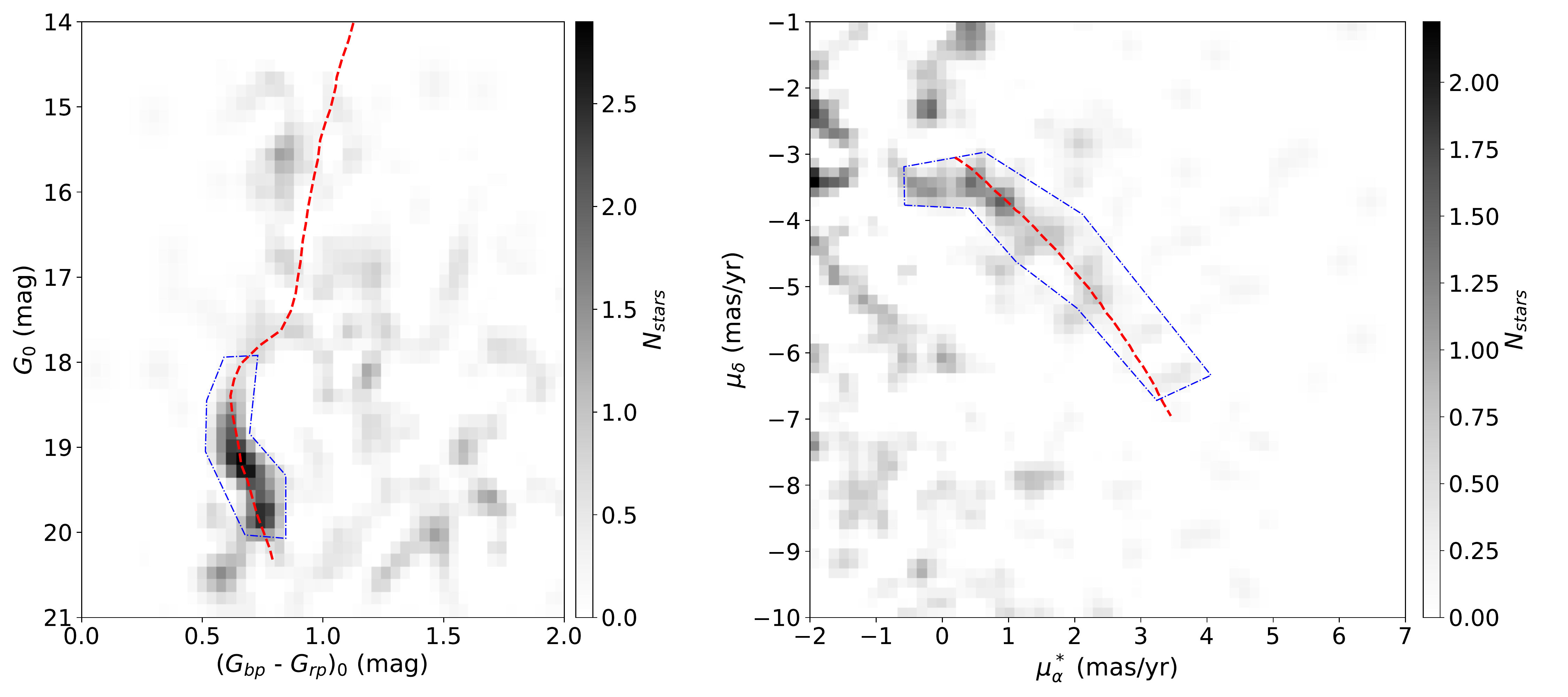}
	\caption{The left panel is a 2D histogram of stars in CMD with PMs selected and background subtracted. The red dashed line represents the CMD-filter moved upward by 0.5 mag. The right panel is a 2D histogram of PMs after CMD selection and background subtraction. The red dashed line represents the PM-filters in this space. Both of the diagrams are smoothed with a 2D Gaussian kernel of $\sigma$ = 1 pixel. The blue polygons represent the CMD and PM selections applied to the stream and off-stream regions.
		\label{fig:leading_CMD_PM}}
\end{figure*}

We further display a background-subtracted binned CMD for the leading tail in the left panel of Figure~\ref{fig:leading_CMD_PM}. The stream region is defined as the sky area around the trajectory (Equation~\ref{equation3}) $\pm 0.8\degr$ in Dec. direction, given the derived width of 1.6$\degr$. The background is estimated by averaging two off-stream regions parallel to the stream, which are obtained by moving the stream region along Dec.-axis by $\pm 4\degr$. This can almost eliminate the effect of stars' number gradient. Before the background subtraction, a PM selection is performed to the stream and off-stream regions as illustrated with the blue polygon in the right panel of Figure~\ref{fig:leading_CMD_PM}, which corresponds to the stream's distribution in PM space. We emphasize that this is a subtraction of star numbers, not weighted counts. During doing so, all stars are taken into account and the restriction in $G_0$ magnitude is no longer required because this limitation is only aimed at lowering noises and making the stream look more prominent on the weighted map.

The CMD bin size is 0.04 mag in color and 0.15 mag in $G_0$ magnitude. The diagram is smoothed with a 2D Gaussian kernel of $\sigma$ = 1 pixel. The red dashed line represents the CMD-filter which is placed on brighter side by 0.5 mag (8 kpc). We set a lower limit of $N_{\rm stars}$ = 0 to better show the overdensities. After the PM selection and background subtraction, the stream's main sequence is obvious and has a good match with the CMD-filter. 

In the right panel of Figure~\ref{fig:leading_CMD_PM}, we present a 2D histogram of PMs. Similarly, before the subtraction between the stream region and the mean of the off-stream regions, a CMD selection is applied to them as shown with the blue polygon in the left panel. The diagram has bin size = 0.15 mas/yr and is also smoothed using a 2D Gaussian with $\sigma$ = 1 pixel. The PM-filters within the range of 210$\degr$ $<$ R.A. $<$ 248$\degr$ are translated into the red dashed line, and around it are some overdensities of stars from the leading tail. We note that the PM-filters from the model stream are good approximations to PMs of the real leading tail. 

We further estimate the stream's surface density and brightness. There are 190 stars within the PM polygon after the background subtraction. This serves as an estimate to the number of the stream stars located in a $35\degr \times 1.6\degr$ region. Thus the surface density is roughly 3.4 stars degree$^{-2}$. The most likely members of the stream can be selected based on the sorting of weights \citep{2019ApJ...884..174G}. We adopt stars in the stream region with weights $>$ 0.91 as the member candidates because the criterion leaves us 190 stars as well. After their $G_0$ fluxes are summed, divided by total area, and transformed into magnitudes, we get a surface brightness of the leading tail to be $\Sigma_G \simeq$ 35.5 mag arcsec$^{-2}$. 

\begin{figure*}
	\includegraphics[width=\linewidth]{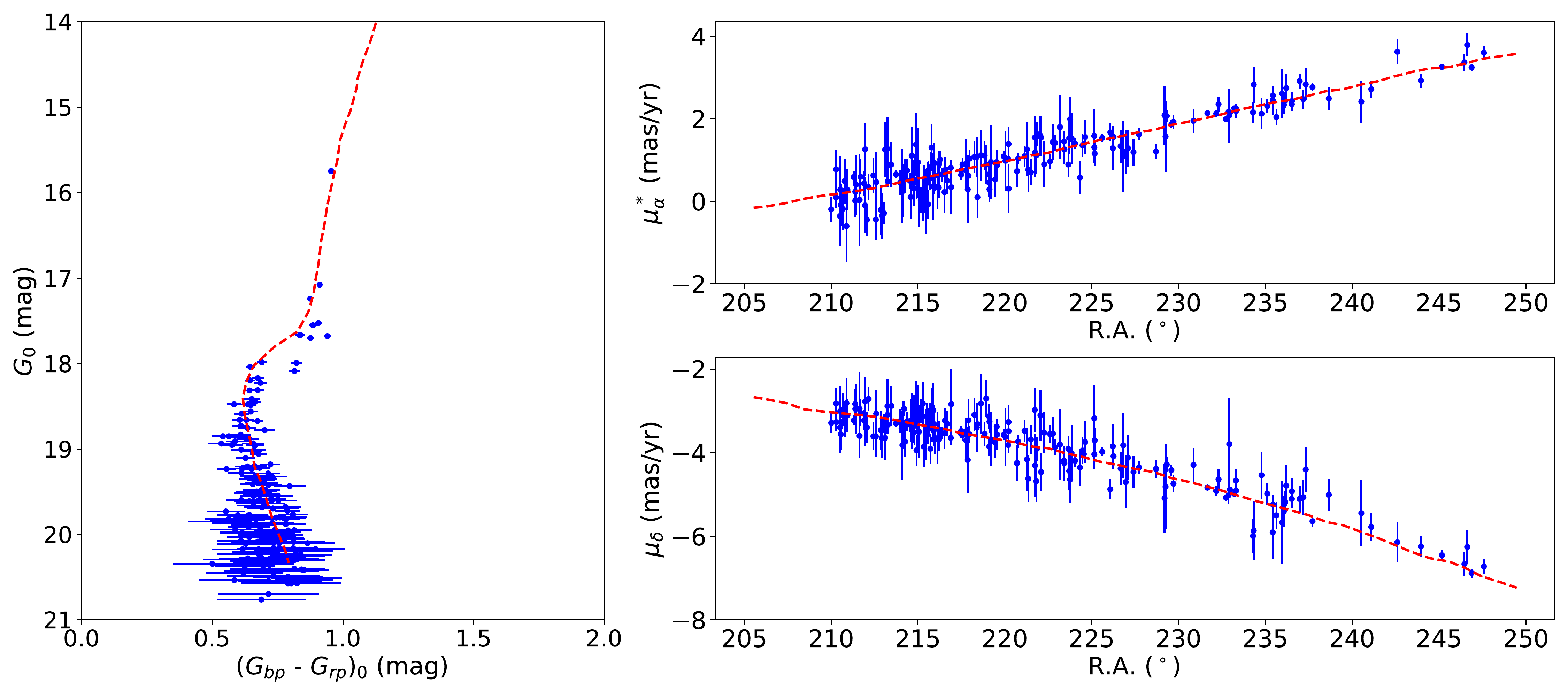}
	\caption{Blue points show candidate stars of the leading tail with weights $>$ 0.91. The red dashed line of the left panel represents the CMD-filter moved upward by 0.5 mag (same as Figure~\ref{fig:leading_CMD_PM}). The red dashed lines of right panels represent the PM-filters. 
		\label{fig:leading_stars}}
\end{figure*}

In Figure~\ref{fig:leading_stars}, we display candidate stars with weights $>$ 0.91 in blue points and compare them to the CMD- and PM-filters shown with red dashed lines. Of course, they are in good agreements because stars are weighted by the filters and higher weights stand for better match. As stated before, three types of $w_{\rm CMD}$ corresponding to three distances are coadded. Here the CMD-filter is placed at a distance of 8 kpc same as Figure~\ref{fig:leading_CMD_PM} and some fainter stars below $G_0$ = 20 mag are matched when the filter is at 10.18 kpc of M3 itself. 

\subsection{The Trailing Tail}

\begin{figure*}
	\centering
	\includegraphics[width=0.9\linewidth]{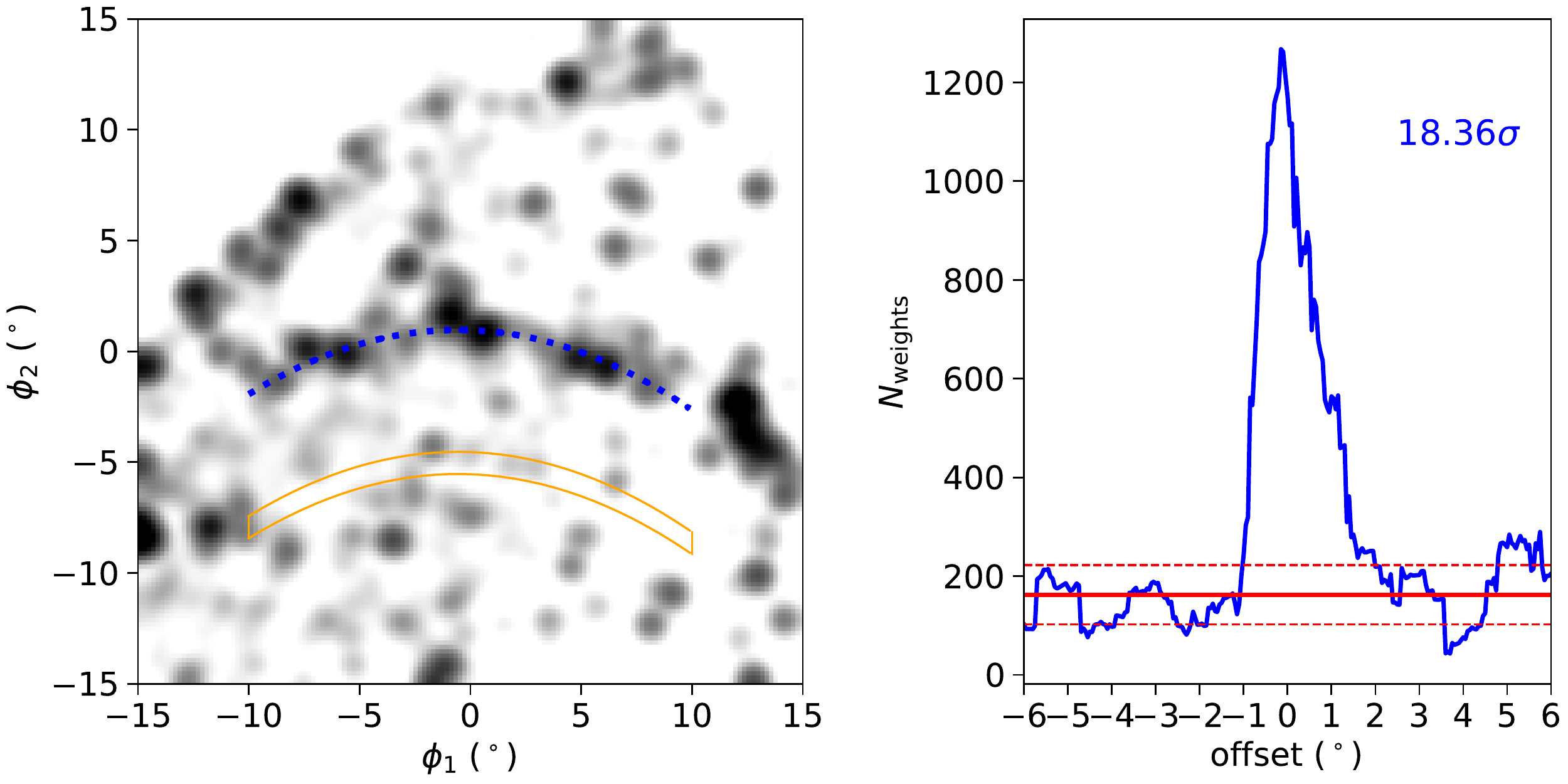}
	\includegraphics[width=0.9\linewidth]{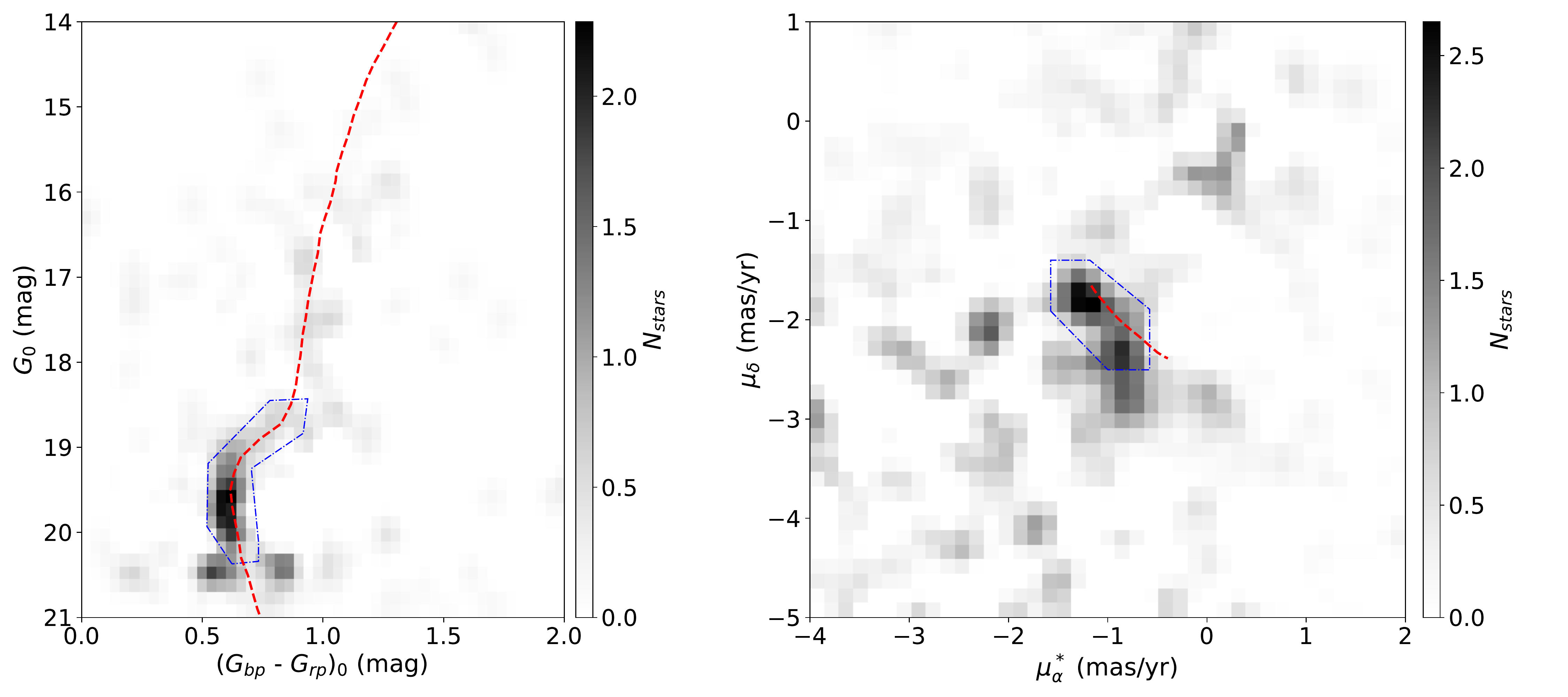}
	\includegraphics[width=0.9\linewidth]{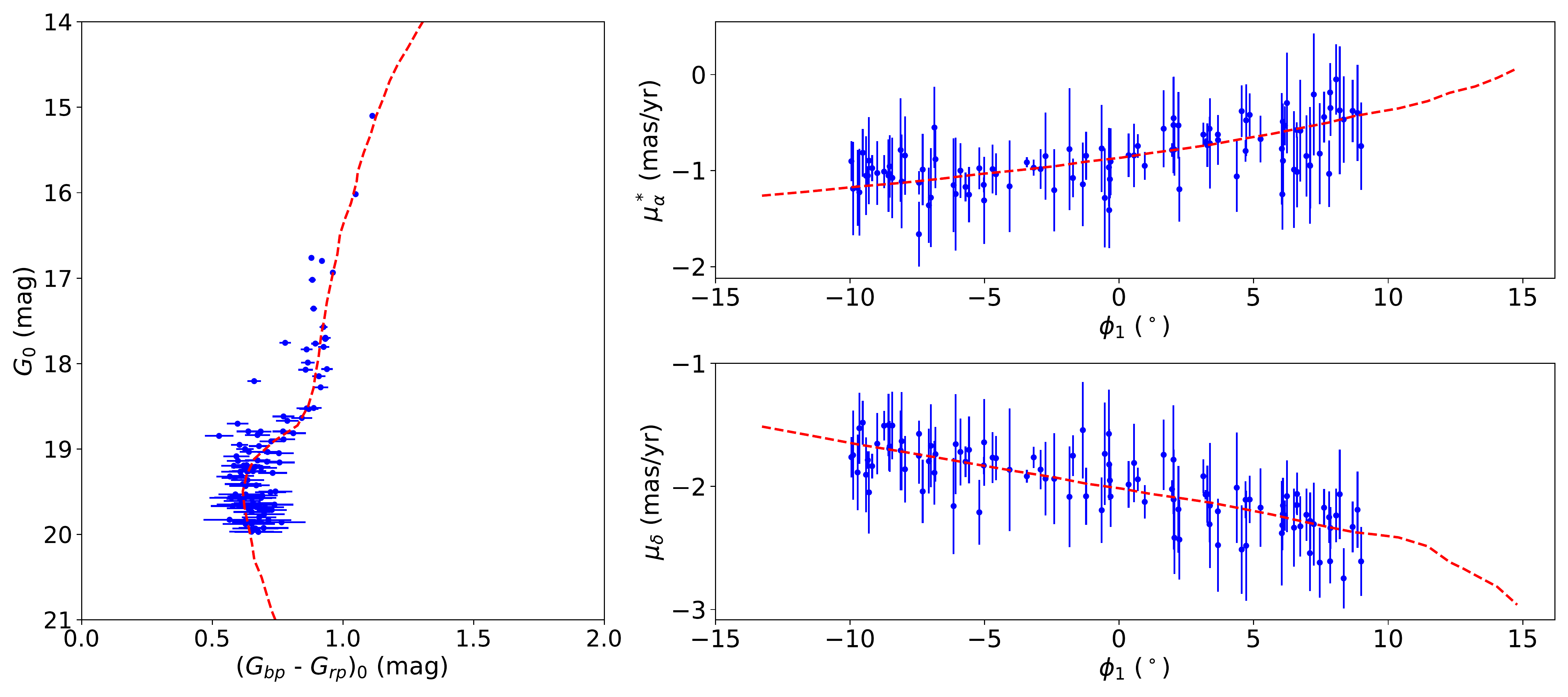}
	\caption{Same analysis as the leading tail, but now for the trailing tail. Top: the weighted map and lateral profile. Middle: stream overdensities in CMD and PM sapce. Bottom: adopted member candidate stars compared to the CMD- and PM-filters. 
		\label{fig:trailing}}
\end{figure*}

In this section, we turn to the trailing tail. We display its zoom-in weighted map on the top left panel in Figure~\ref{fig:trailing}. For a clearer illustration, the map is plotted in a stream-aligned coordinate frame, which is a rotation from ICRS based on two endpoints (R.A., Dec.) = (186.45, 19.06)$\degr$ and (197.20, 27.76)$\degr$. Such a rotation can be easily done by some tools, e.g., \texttt{Gala}\footnote{\url{https://gala.adrian.pw/en/latest/coordinates/greatcircle.html\#creating-a-coordinate-frame-from-two-points-along-a-great-circle}.} \citep{2017JOSS....2..388P}. It is worth noting that there are two very strong stream signals at R.A. $\simeq$ 198$\degr$ in Figure~\ref{fig:weighted_map} which is because two stars (\texttt{source\_id} = 1460891108071403008 and 1462653522131570816) are assigned big weights and dramatically increase total weights of corresponding pixels. Actually, we can consider these two stars as member candidates, but we exclude them in top panels of Figure~\ref{fig:trailing} because the lateral profile will be heavily affected by them. Note that we are just decreasing the stream's strength as appropriate and background noises are not changed at all. 

In the top left panel of Figure~\ref{fig:trailing}, the trailing tail is fitted with a second-order polynomial of
\begin{equation}
	\phi_2 = -0.03250747 \phi_1^2 -0.03606698 \phi_1 + 0.95664055
\end{equation}
with -10$\degr$ $< \phi_1 <$ 10$\degr$. In the ICRS frame, the trajectory can be described with a third-order polynomial:
\begin{equation}
	\rm{Dec.} = 3.45473581 \times 10^{-3} \times \rm{R.A.}^3 -2.06257886 \times \rm{R.A.}^2 + 4.10633299 \times 10^2 \times \rm{R.A.} -2.72338774 \times 10^4
\end{equation}
where 185$\degr$ $<$ R.A. $<$ 202$\degr$. Similarly, moving the 1$\degr$-wide mask across the stream will create its lateral profile as displayed in the top right panel. The significance is 18.36$\sigma$ above the background, and from the profile we find its FWHM to be about 1.4$\degr$.

The background-subtracted binned CMD and PM space are presented in middle panels of Figure~\ref{fig:trailing}. The way we create it is exactly the same as that applied on the leading tail. Here the CMD-filter is moved downward by 0.6 mag and the PM-filters are in the range 185$\degr$ $<$ R.A. $<$ 202$\degr$ covering the detected portion. Although highly contaminated by field stars for the trailing part, overdensities resembling the filters still appear, and provide evidence of its existence. There are 100 stars within the PM polygon after subtracting background which is an estimated number of the trailing tail stars. The stream region we defined here is 1.4$\degr$ wide and 21$\degr$ long. Thus the number density is about 3.4 stars degree$^{-2}$. We adopt stars in the stream region with weights $>$ 5.3 as member candidates because such a threshold also leaves us 100 stars, and a surface brightness of $\Sigma_G \simeq$ 34.8 mag arcsec$^{-2}$ is found for the trailing tail. The measurements are affected by the candidate number which is further related to how many stars there are in the PM polygon for the stream and off-stream regions. The latter quantities follow Poisson statistics that will lead to a variation on the order of $\sim$ 1 star degree$^{-2}$ in density and $\sim$ 0.1 mag arcsec$^{-2}$ in brightness. The same is true for the leading tail. 

Finally in bottom panels of Figure~\ref{fig:trailing}, we show adopted members with weights $>$ 5.3 in blue points and the CMD- and PM-filters with red dashed lines. The CMD-filter is placed at 13.5 kpc same as the middle left panel. It is noted that there are few stars fainter than $G_0$ = 20 mag since a relative high weight threshold is used here.

\subsection{Distance along the stream}

Although 290 candidate members of M3 tidal tails are selected, the uncertainties of parallaxes from Gaia DR3 are not good enough to derive valid distances. To inspect the distance range covered by the stream, we do some estimates based on a CMD fitting. 

\begin{figure}
	\includegraphics[width=\linewidth]{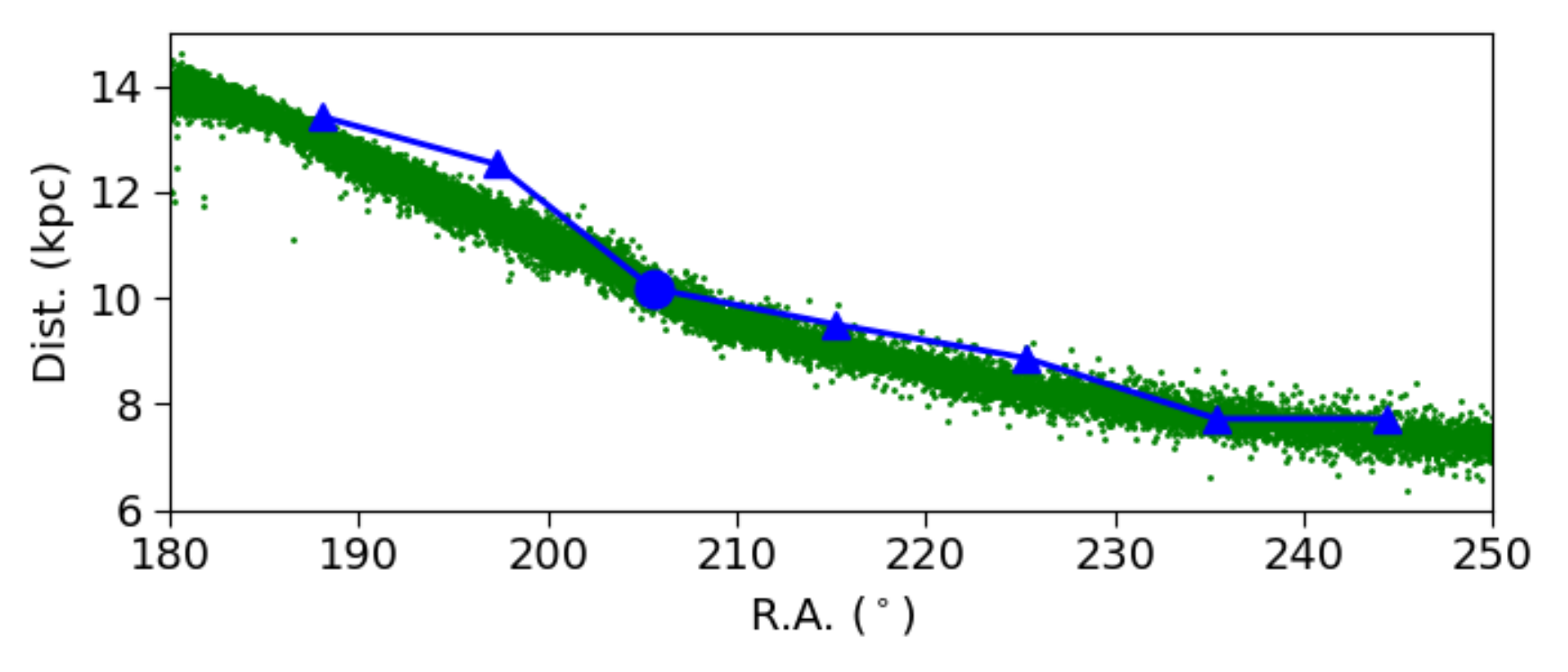}
	\caption{Heliocentric distance variation along the stream. Green dots represent the model stream. Blue triangles show our estimates on the distance of each stream segment in a certain R.A. range. The blue circle indicates M3.
	   \label{fig:distance}}
\end{figure}

From the lateral profiles, it can be deduced that weights of stream regions (|offset| $< 0.8\degr$ for the leading tail and |offset| $< 0.7\degr$ for the trailing tail) are dominated by stream stars rather than the background. Hence we can cut the stream into several parts in different R.A. ranges and use total weights of each part to measure matching degree between stream stars and the CMD-filter. Specifically, the trailing tail is divided into two segments in R.A. ($\degr$) range [184, 193] and [193, 203], and the leading one consists of four divisions in R.A. range [210, 220], [220, 230], [230, 240], and [240, 250]. Each section is considered to have a mean distance and stars are given weights using the CMD-filter at different distance modulus (dm) meanwhile the PM-filters remain the same as before. For the leading tail on closer side, variation of dm is $\Delta$dm = [-1, 0, 0.05] ([start, stop, step]) mag, while for the trailing farther side $\Delta$dm is [0, 1, 0.05] mag. The best fitted distance corresponds to the largest sum of weights. Results are shown in Figure~\ref{fig:distance}, from which the inferred distances (triangles) generally agree with the model stream.

\subsection{Sv\"{o}l as M3's leading tail}

\begin{figure*}
	\includegraphics[width=0.36\linewidth]{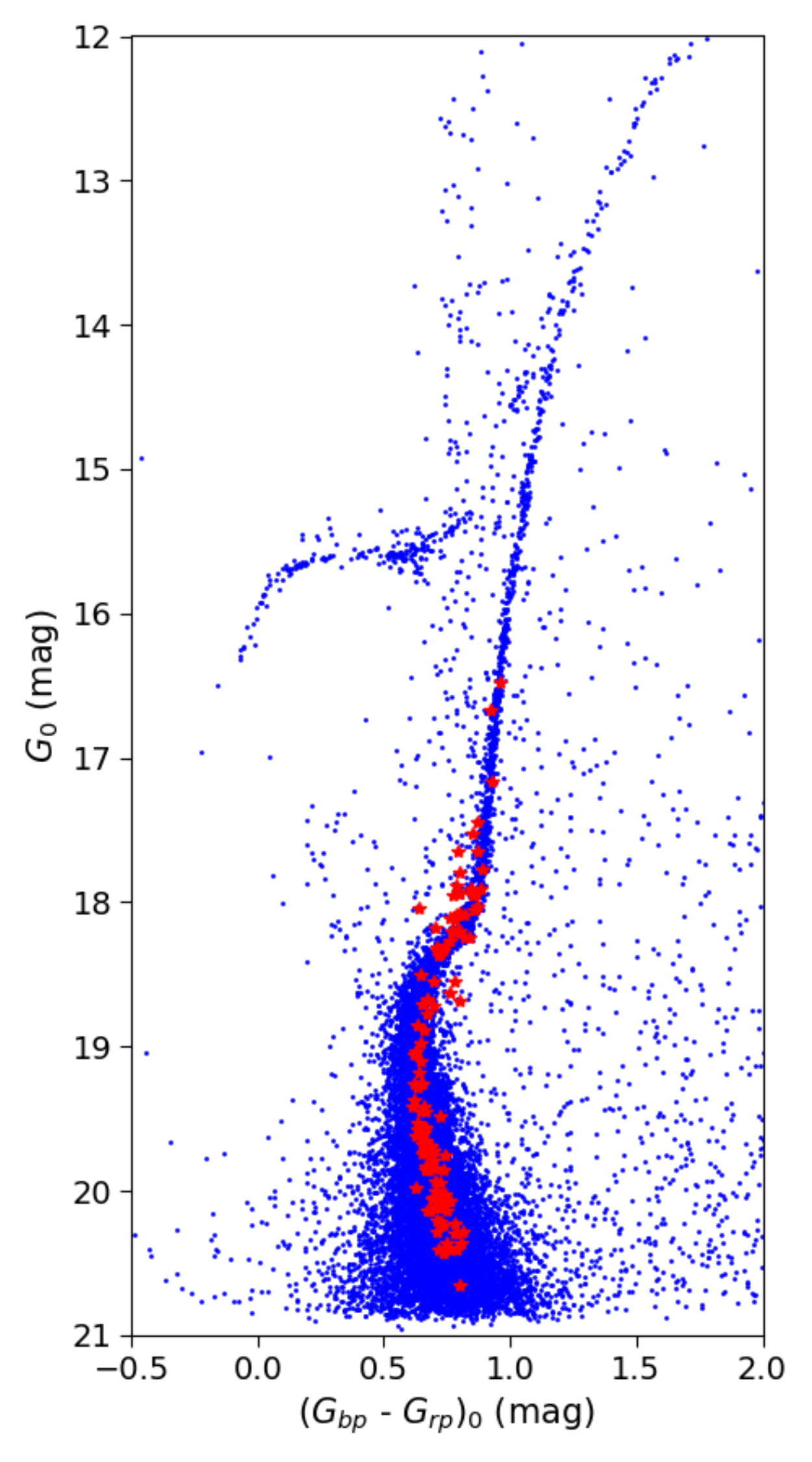}
	\includegraphics[width=0.64\linewidth]{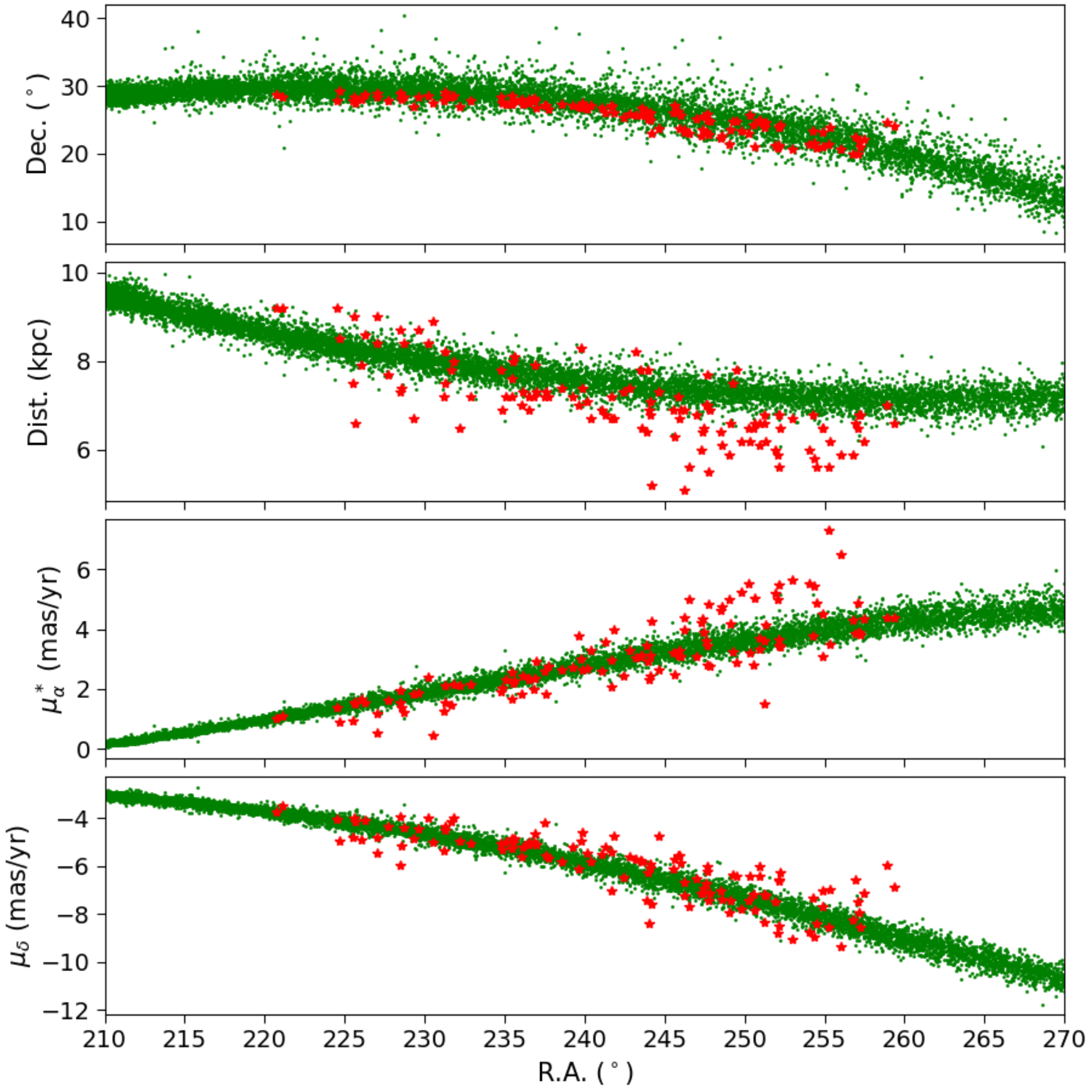}
	\caption{The left panel is a CMD of M3 (blue dots) and Sv\"{o}l (red stars). The stream stars are all placed at M3's distance. The right panel shows the model stream (green dots) and Sv\"{o}l (red stars) in planes of Dec., heliocentric distance, $\mu^*_{\alpha}$ and $\mu_{\delta}$ as a function of R.A., from top to bottom. 
		\label{fig:comparison}}
\end{figure*}

\citet{2021ApJ...909L..26B} has already associated Sv\"{o}l \citep{2021ApJ...914..123I} with M3 based on their energies and angular momenta. Here we present a comparison of them as an elaboration. In Figure~\ref{fig:comparison}, the model particles, stars within M3's tidal radius and Sv\"{o}l are plotted in green dots, blue dots and red stars, respectively. Sv\"{o}l stars' distances are estimated by \texttt{STREAMFINDER} \citep{2018MNRAS.477.4063M} and they are all placed at M3's distance in the CMD panel. Good consistency can be seen from both panels and results of \citet{2021ApJ...909L..26B} are verified, which is that M3 is the progenitor of Sv\"{o}l.

\subsection{Spectroscopic stream member stars}

\begin{figure}
	\includegraphics[width=\linewidth]{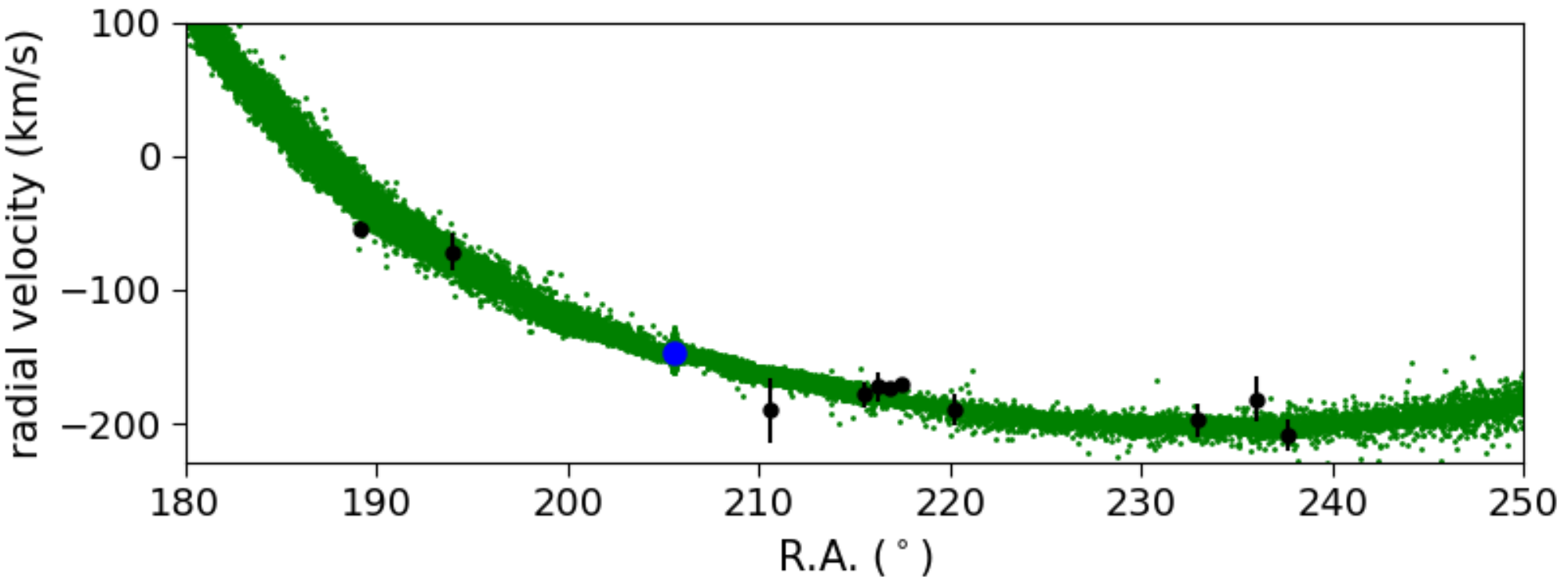}
	\caption{The green dots show the model stream's radial velocity as a function of R.A.. The member stars in Table~\ref{tab:members} are plotted in the black points. The blue circle stands for M3.
		\label{fig:RV}}
\end{figure}

As mentioned before, we treat stars with weights $>$ 0.91 in leading stream region and $>$ 5.3 in trailing stream region as member candidates given that these thresholds leave us star numbers identical to those estimated from subtraction of stars in stream and off-stream regions. We collect spectroscopic data from LAMOST DR8 \citep{2012RAA....12.1197C,2012RAA....12..723Z} and SDSS/SEGUE DR9 \citep{2012ApJS..203...21A}, and cross-match them with the candidates as well as Sv\"{o}l stars. After discarding outliers inconsistent with the model stream's radial velocity trend along R.A., we get a total of 11 stream stars, out of which 10 are from our candidates and one is supplemented by Sv\"{o}l. They are tabulated in Table~\ref{tab:members} and their radial velocities as a function of R.A. are presented in Figure~\ref{fig:RV}. As a validation, their [Fe/H] are centered at -1.5 dex of M3 \citep[][2010 edition]{1996AJ....112.1487H} although we do not restrict their metallicities at all. 

\begin{table*}
	\centering
	\caption{Member stars of M3' tidal tails. Column 1 to 5 are from Gaia DR3. Column 6 and 9 are from spectroscopic surveys as said in the last column. Column 7 and 8 are extinction-corrected magnitude and color. The penultimate star is complemented from Sv\"{o}l stars.}
	\label{tab:members}
	\begin{tabular}{cccccccccl} 
		\hline
		source\_id & R.A. & Dec. & $\mu^*_{\alpha}$ & $\mu_{\delta}$ & $V_r$ & $G_0$ & $(BP-RP)_0$ & [Fe/H] & Survey \\
		  & ($\degr$) & ($\degr$) & (mas yr$^{-1}$) & (mas yr$^{-1}$) & (km s$^{-1}$) & (mag) & (mag) & (dex) &  \\
		\hline
		3958917028852911232 & 189.1316 & 23.4988 & -1.0765 & -1.7502 & -54.61$\pm$5.45 & 18.52 & 0.89 & -1.590$\pm$0.080 & SEGUE\\
		3958529828960889472 & 193.9868 & 26.1820 & -0.5636 & -2.3073 & -71.98$\pm$14.07 & 17.76 & 0.78 & -1.481$\pm$0.325 & LAMOST\\
		1452515161633128576 & 210.6023 & 28.2220 & 0.1981 & -3.0312 & -190.03$\pm$24.19 & 17.55 & 0.89 & -1.659$\pm$0.381 & LAMOST\\
		1259791217328624896 & 215.5418 & 27.9655 & 0.7228 & -3.5630 & -178.36$\pm$9.13 & 18.85 & 0.56 & -1.786$\pm$0.047 & SEGUE\\
		1284032803020905344 & 216.2726 & 29.2915 & 1.0164 & -3.5633 & -172.65$\pm$11.08 & 18.94 & 0.53 & -1.548$\pm$0.025 & SEGUE\\
		1283742394512311424 & 216.8810 & 28.4759 & 0.8161 & -3.6425 & -174.22$\pm$5.67 & 18.66 & 0.63 & -1.392$\pm$0.070 & SEGUE\\
		1283722775101748480 & 217.4780 & 28.6154 & 0.6439 & -3.4800 & -170.35$\pm$3.80 & 17.98 & 0.69 & -1.748$\pm$0.029 & SEGUE\\
		1280001688451645312 & 220.1860 & 27.7717 & 1.0302 & -3.8147 & -189.88$\pm$11.44 & 17.08 & 0.91 & -1.650$\pm$0.276 & LAMOST\\
		1270836846719121536 & 232.9526 & 27.6245 & 2.1492 & -4.8814 & -197.73$\pm$12.42 & 17.24 & 0.87 & -1.331$\pm$0.253 & LAMOST\\
		1224502013678132736 & 236.0643 & 27.9842 & 1.8500 & -5.5970 & -182.08$\pm$16.89 & 17.36 & 0.70 & -1.870$\pm$0.136 & LAMOST\\
		1223324234863007616 & 237.7107 & 26.5518 & 2.7691 & -5.6384 & -209.36$\pm$11.73 & 18.49 & 0.65 & -1.357$\pm$0.073 & SEGUE\\
		\hline
	\end{tabular}
\end{table*}

\section{Conclusion} \label{sec:summary}

In this work, we present a detailed analysis on tidal tails of M3. We first introduce a revised matched filter that assigns weights through a ratio of the target to background in CMD and PM space simultaneously. Some clear extra-tidal structures are revealed in the surrounding area of M3 and it motivates us to search for the existence of long tidal tails of the cluster. 

We then produce the cluster's model stream and use it as a prediction of how PMs of tails change along spatial coordinates. The locus of M3 in the CMD is further employed to assign weights. By utilizing such a modified matched filter \citep{2019ApJ...884..174G}, we successfully detect M3's leading and trailing tails. 

We give deeper insights to the both tails. The leading tail covers a length of 35$\degr$ and its lateral profile indicates a width of 1.6$\degr$. The background-subtracted CMD and PM space show clear overdensities of the tail stars which follow distributions of the filters we used. The surface density and brightness are about 3.4 stars degree$^{-2}$ and 35.5 mag arcsec$^{-2}$, respectively. As to the trailing tail, the detected portion is 21$\degr$ long and 1.4$\degr$ wide on average. In CMD and PM panels also appear star overdensities. The number density is roughly 3.4 stars degree$^{-2}$ and surface brightness is 34.8 mag arcsec$^{-2}$. By applying a CMD fitting to different stream divisions, a distance variation along the tails is obtained which resembles the model prediction. There is good consistency between known Sv\"{o}l stream and M3 model particles in phase space, and Sv\"{o}l also matches with M3 itself in CMD. The results support that M3 is the mother cluster of Sv\"{o}l. At last, we conclude 11 member stars of M3 tidal stream for which there are spectroscopic data.

\begin{acknowledgments}
	We thank the referee for the thorough reviews that helped us to improve the manuscript. Y.Y. thanks Ying-Hua Zhang for her kind help concerning the matched filter analysis. This study is supported by the National Natural Science Foundation of China under grant nos 11988101, 12273055, 11973048, 11927804, 11890694 and 12261141689, and the National Key R\&D Program of China, grant no. 2019YFA0405500.  
	
	Guoshoujing Telescope (the Large Sky Area Multi-Object Fiber Spectroscopic Telescope LAMOST) is a National Major Scientific Project built by the Chinese Academy of Sciences. Funding for the project has been provided by the National Development and Reform Commission. LAMOST is operated and managed by the National Astronomical Observatories, Chinese Academy of Sciences.
	
	This work presents results from the European Space Agency (ESA) space mission $Gaia$. $Gaia$ data are being processed by the $Gaia$ Data Processing and Analysis Consortium (DPAC). Funding for the DPAC is provided by national institutions, in particular the institutions participating in the $Gaia$ MultiLateral Agreement (MLA). The $Gaia$ mission website is \url{https://www.cosmos.esa.int/gaia}. The $Gaia$ archive website is \url{https://archives.esac.esa.int/gaia}.
	
	Funding for the Sloan Digital Sky Survey IV has been provided by the Alfred P. Sloan Foundation, the U.S. Department of Energy Office of Science, and the Participating Institutions. 
	
	SDSS-IV acknowledges support and resources from the Center for High Performance Computing  at the University of Utah. The SDSS website is \url{www.sdss.org}.
	
	SDSS-IV is managed by the Astrophysical Research Consortium for the Participating Institutions of the SDSS Collaboration including the Brazilian Participation Group, the Carnegie Institution for Science, Carnegie Mellon University, Center for Astrophysics | Harvard \& Smithsonian, the Chilean Participation Group, the French Participation Group, Instituto de Astrof\'isica de Canarias, The Johns Hopkins University, Kavli Institute for the Physics and Mathematics of the Universe (IPMU) / University of Tokyo, the Korean Participation Group, Lawrence Berkeley National Laboratory, Leibniz Institut f\"ur Astrophysik Potsdam (AIP),  Max-Planck-Institut f\"ur Astronomie (MPIA Heidelberg), Max-Planck-Institut f\"ur Astrophysik (MPA Garching), Max-Planck-Institut f\"ur Extraterrestrische Physik (MPE), National Astronomical Observatories of China, New Mexico State University, New York University, University of Notre Dame, Observat\'ario Nacional / MCTI, The Ohio State University, Pennsylvania State University, Shanghai Astronomical Observatory, United Kingdom Participation Group, Universidad Nacional Aut\'onoma de M\'exico, University of Arizona, University of Colorado Boulder, University of Oxford, University of Portsmouth, University of Utah, University of Virginia, University of Washington, University of Wisconsin, Vanderbilt University, and Yale University.
\end{acknowledgments}

%


\appendix

\section{Comparing the Trailing Tail to the Field Stars} \label{appendix}

\begin{figure*}
	\includegraphics[width=\linewidth]{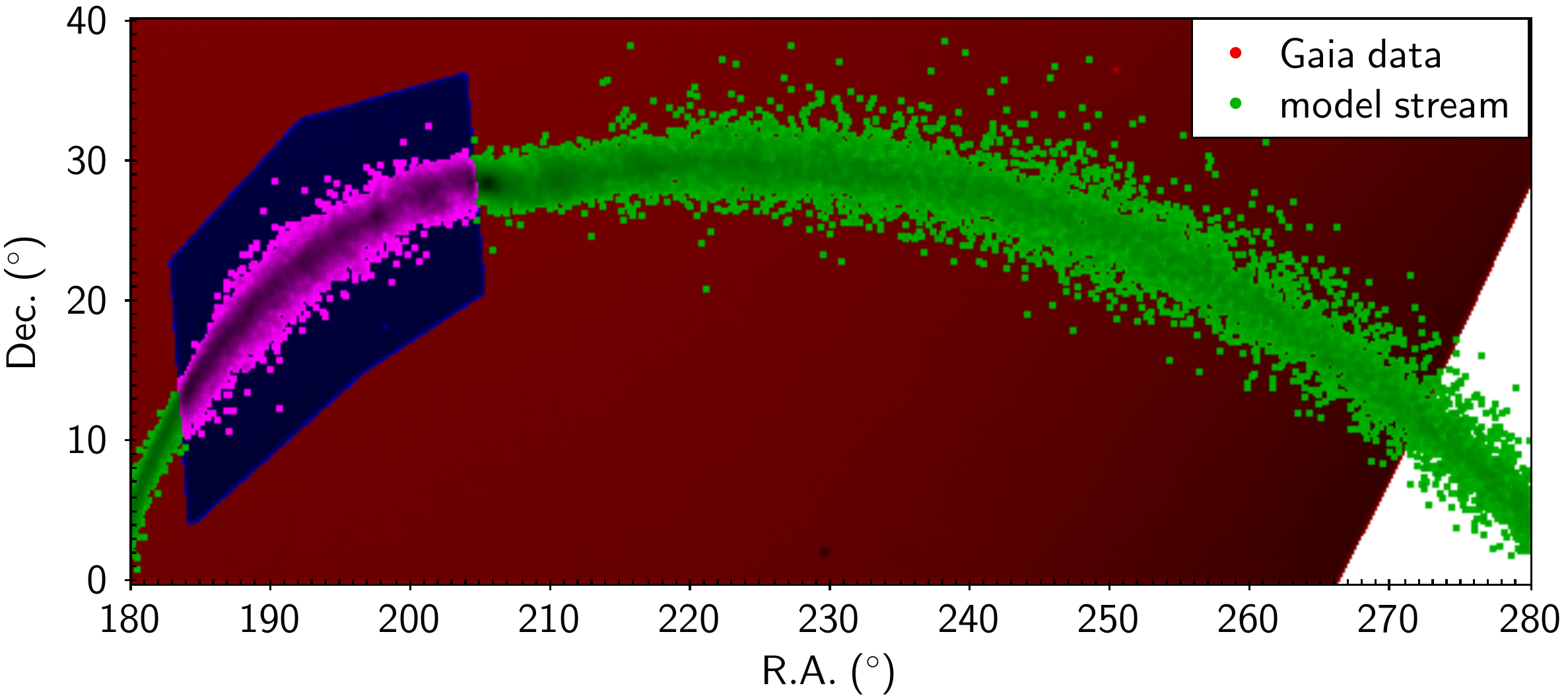}
	\includegraphics[width=0.55\linewidth]{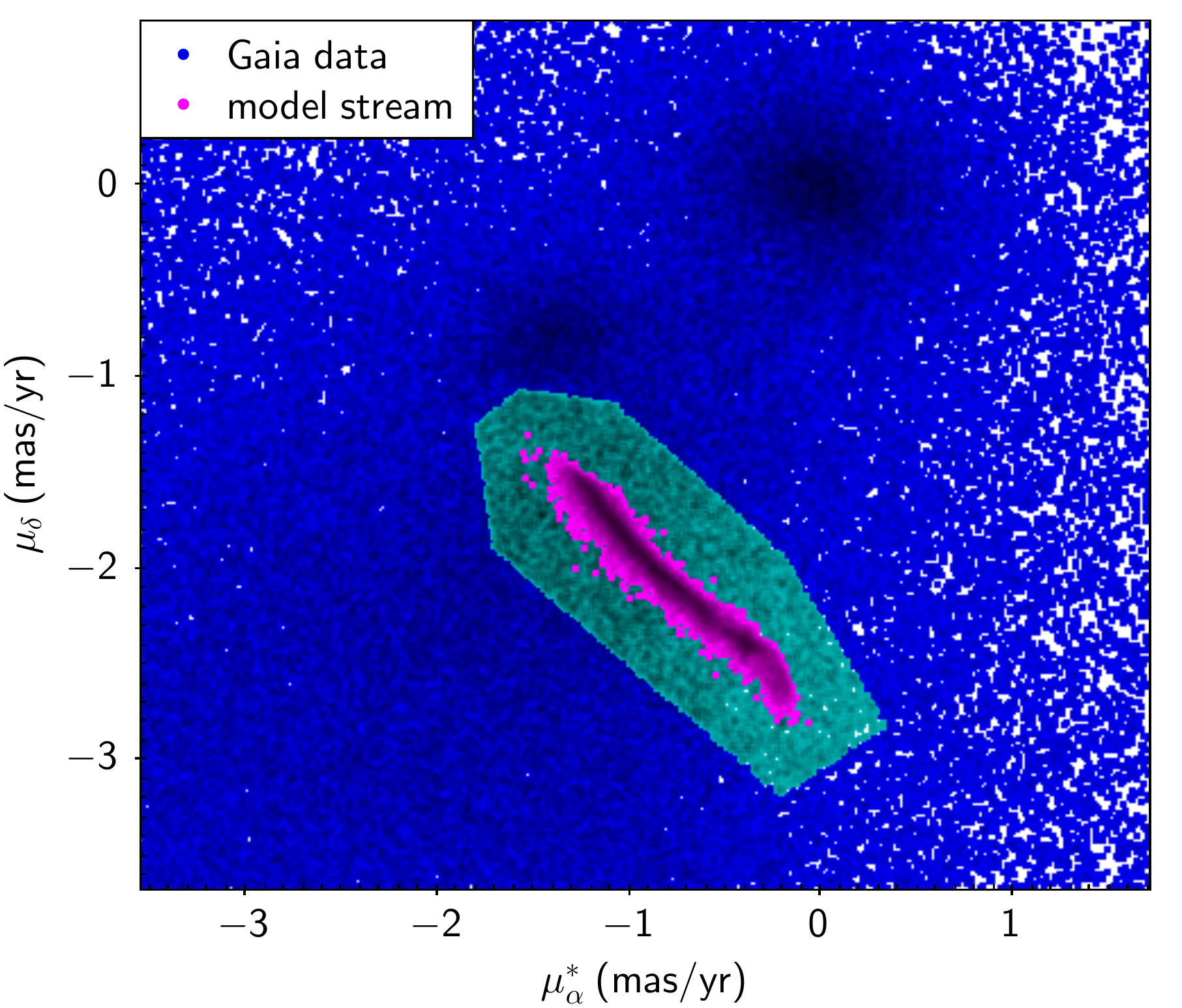}
	\includegraphics[width=0.45\linewidth]{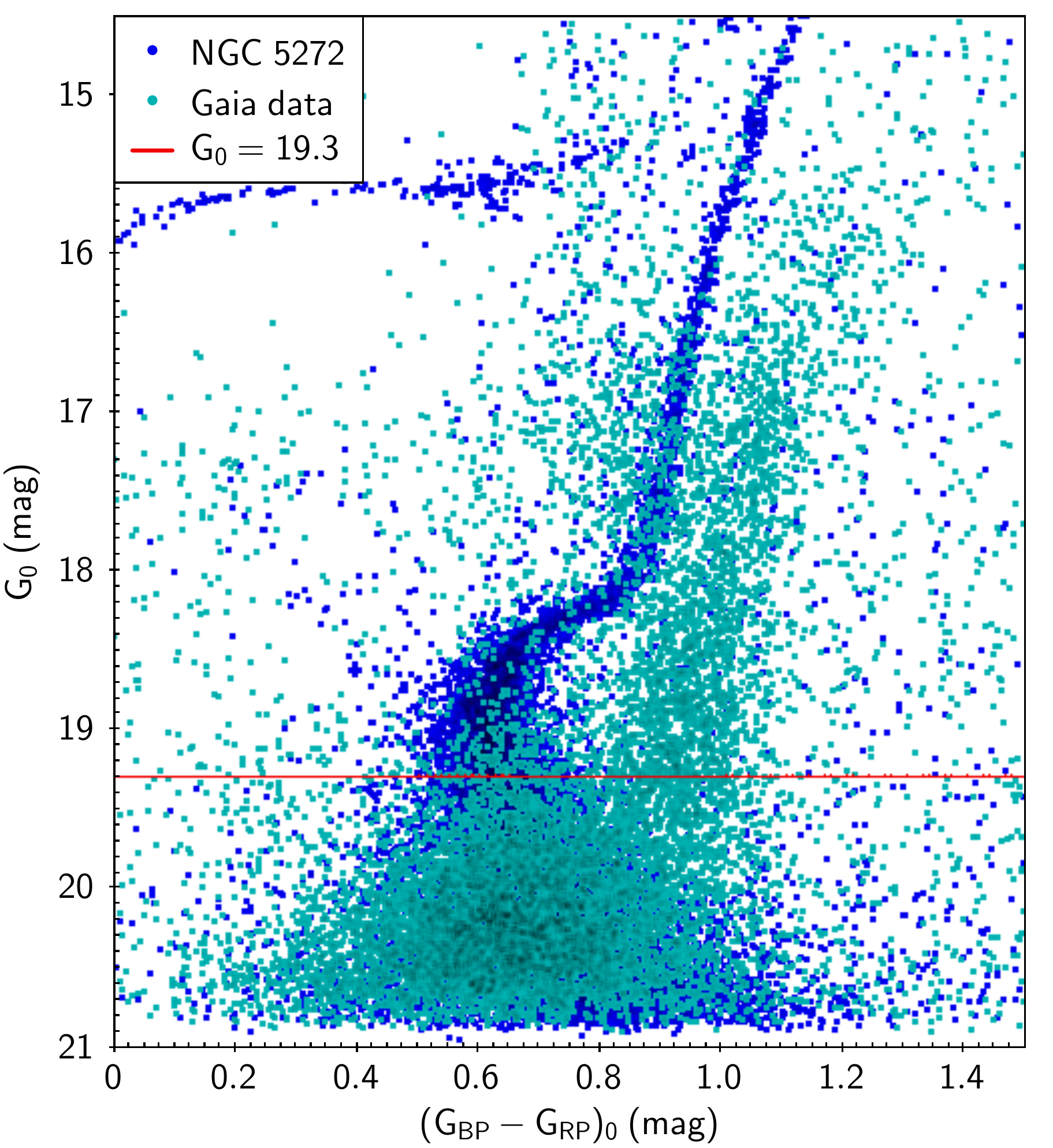}
	\caption{An illustration of how come we set $G_0 <$ 19.3 mag in Figure~\ref{fig:weighted_map}. Based on the model, we select nearby stars on sky (blue dots in upper panel) and further in PM space (cyan dots in lower left panel). These stars are compared to M3 on CMD (lower right panel) and it can be found that $G_0 <$ 19.3 mag is able to discard most of contaminants.
		\label{fig:appendix}}
\end{figure*}

We give a simple illustration of why we require $G_0 <$ 19.3 mag in Section~\ref{sec:awholelook} to distinguish the trailing tail from field stars. In the upper panel of Figure~\ref{fig:appendix}, we first select stars (blue) in a neighbouring area based on the model. These stars are plotted in PM space in lower left panel and the counterpart of the model stream is shown in pink. The surrounding of the model is full of stars which means that we can hardly tell the stream from field stars just using PMs. Considering that PM-filters will highlight nearby stars, we directly select them (cyan) and plot them on a CMD in lower right panel. It can be found that a majority of main sequence stars of M3 (blue) is still highly contaminated. This is why we can not see the trailing tail clearly in upper panel of Figure~\ref{fig:weighted_map}. However, there are fewer contaminants overlapping with the cluster if we only focus on stars brighter than $G_0$ = 19.3 mag. As a result, most of surrounding field signals disappear in lower panel of Figure~\ref{fig:weighted_map} when we further get rid of fainter stars.

\bibliography{sample631}{}
\bibliographystyle{aasjournal}



\end{document}